\documentclass[twoside,twocolumn]{article}

\usepackage{reflector}
\usepackage{named}
\usepackage{float}
\usepackage{caption}
\renewcommand{\shortcite}[1]{}

\usepackage[ruled]{algorithm2e} 

\SetAlFnt{\small}
\SetAlCapFnt{\small}
\SetAlCapNameFnt{\small}
\SetAlCapHSkip{0pt}
\IncMargin{-\parindent}

\usepackage[T1]{fontenc} 

\usepackage[english]{babel} 

\usepackage[hmargin={1cm, 1cm},top=32mm,columnsep=20pt]{geometry} 
\usepackage{booktabs} 

\usepackage{enumitem} 
\setlist[itemize]{noitemsep} 

\usepackage{abstract} 

\usepackage{titlesec} 
\titleformat{\section}[block]{\large\scshape\centering}{\thesection.}{1em}{} 
\titleformat{\subsection}[block]{\large}{\thesubsection.}{1em}{} 

\usepackage{fancyhdr} 
\usepackage{titling} 

\usepackage{hyperref} 

\pretitle{\begin{center}\Huge\bfseries} 
\posttitle{\end{center}} 
\title{Light in Power: A General and Parameter-free Algorithm for Caustic Design} 
\author{%
\textsc{Jocelyn Meyron} \\[1ex] 
\normalsize Univ. Grenoble Alpes, CNRS, LJK \\ 
\and 
\textsc{Quentin M\'erigot} \\[1ex] 
\normalsize Univ. Paris-Sud, CNRS, LMO \\ 
\and
\textsc{Boris Thibert} \\[1ex] 
\normalsize Univ. Grenoble Alpes, CNRS, LJK \\ 
}
\date{\today} 
\renewcommand{\maketitlehookd}{%
\begin{abstract}
We present in this paper a generic and parameter-free algorithm to efficiently
build a wide variety of optical components, such as mirrors or lenses, that
satisfy some light energy constraints. In all of our problems, one is given a
collimated or point light source and a desired illumination after reflection or
refraction and the goal is to design the geometry of a mirror or lens which
transports exactly the light emitted by the source onto the target. We first
propose a general framework and show that eight different optical component
design problems amount to solving a \textit{light energy conservation} equation
that involves the computation of \textit{visibility diagrams}. We then show that
these diagrams all have the same structure and can be obtained by intersecting a
3D Power diagram with a planar or spherical domain. This allows us to propose an
efficient and fully generic algorithm capable to solve these eight optical
component design problems.  \boris{The support of the prescribed target illumination can be a set of directions or a set of points located at a finite distance.} Our solutions satisfy design constraints such as
convexity or concavity. We show the effectiveness of our algorithm on 
simulated and fabricated examples.
\end{abstract}
    \begin{center}
    \includegraphics[height=3.9cm]{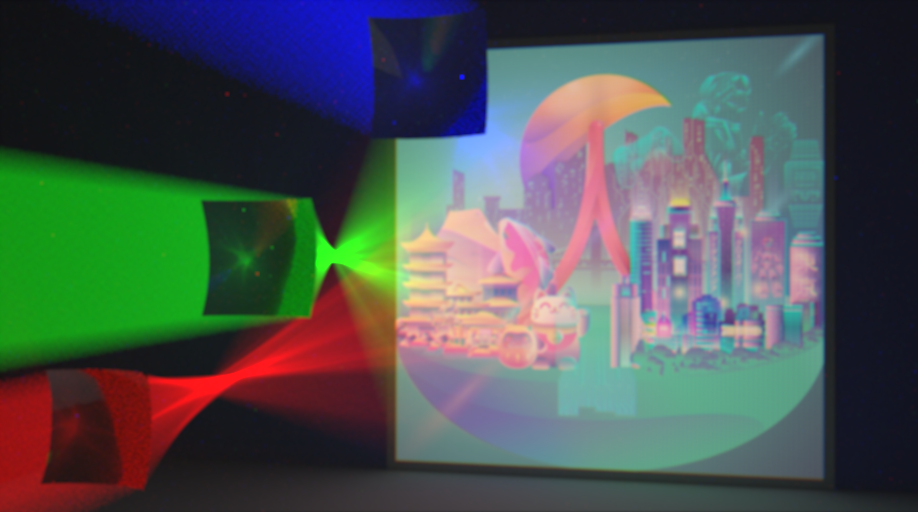}
    \includegraphics[height=3.9cm]{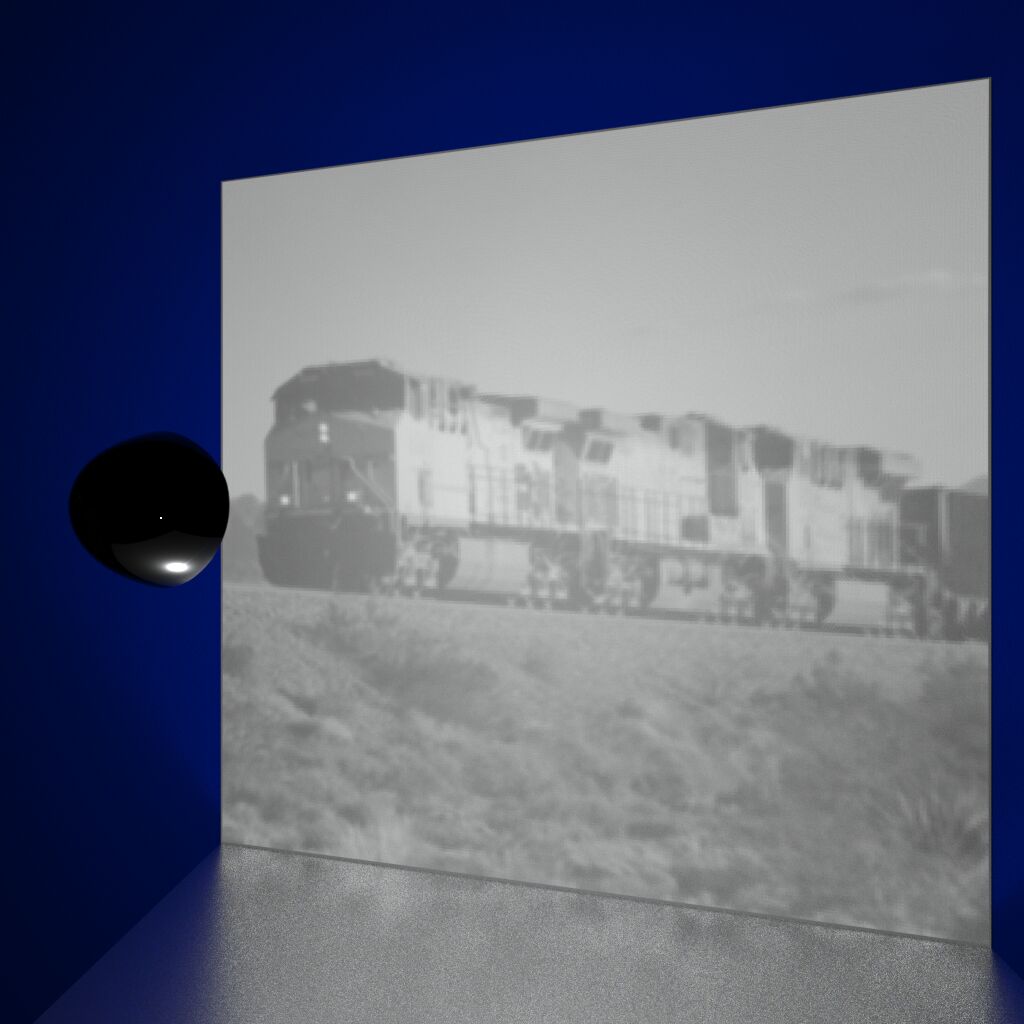} 
    \includegraphics[height=3.9cm]{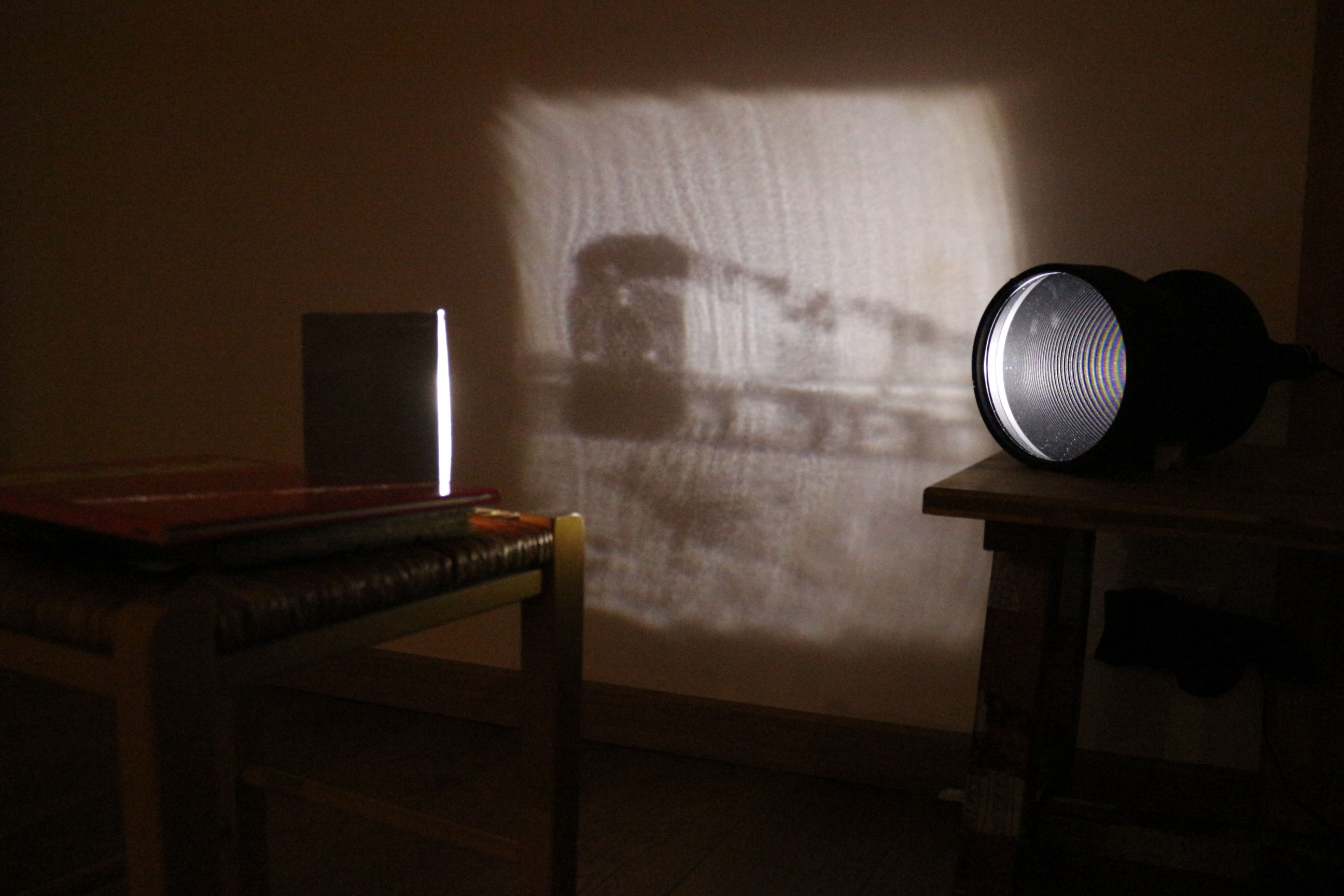}
    \includegraphics[height=3.9cm]{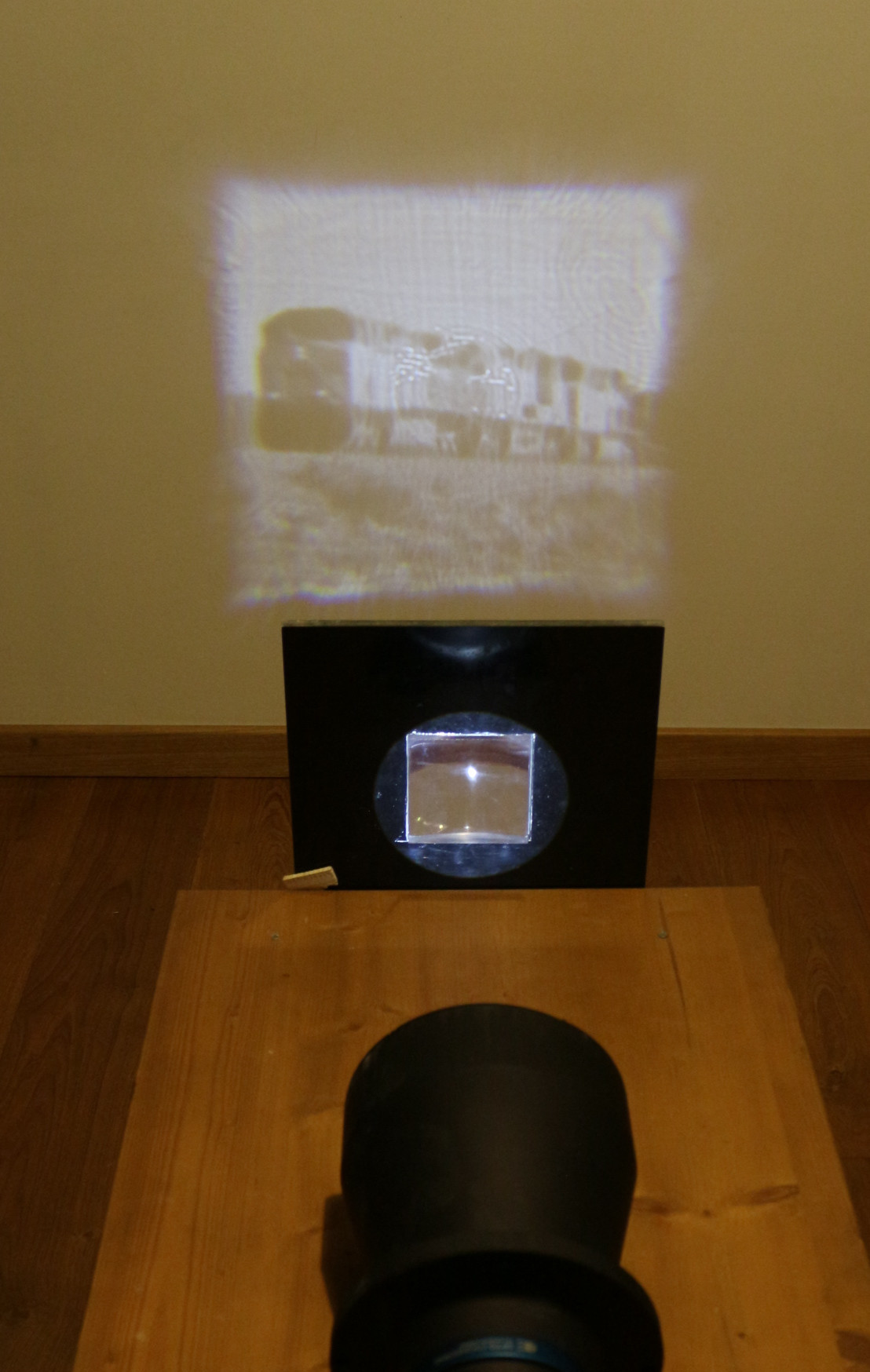}
    \captionof{figure}{Our algorithm can be used to design mirrors and lenses that reflect
        or refract collimated or point light sources onto a prescribed distribution of light. \textit{From left to right:} \jocelyn{Three
            lenses that refract the three channels of a color image; Mirror
            that reflects a point light source (located inside the mirror);
            Fabricated mirror that reflects a collimated light source;
            Fabricated lens that refracts a collimated light source.}}\label{fig:teaser}
    \end{center}
}

\begin{document}

\maketitle

\section{Introduction}

The field of non-imaging optics deals with the design of optical
components whose goal is to transfer the radiation emitted by a light
source onto a prescribed target. This question is at the heart of many
applications where one wants to optimize the use of light energy by
decreasing light loss or light pollution. Such problems appear in the
design of car beams, public lighting, solar ovens and hydroponic
agriculture. This problem has also been considered under the
name of \emph{caustic design}, with applications in architecture and
interior decoration \cite{finckh2010geometry}.

In this paper, we consider the problem of designing a wide variety of
mirrors and lenses that satisfy different kinds of light energy
constraints. To be a little bit more specific, in each problem that we
consider, one is given a light source and a desired illumination after
reflection or refraction which is called the target. The goal is to
design the geometry of a mirror or lens which transports exactly the
light emitted by the source onto the target. The design of such
optical components can be thought of as an \emph{inverse problem},
where the \emph{\jocelynbis{forward} problem} would be the simulation of the target
illumination from the description of the light source and the geometry
of the mirror or lens.

In practice, the mirror or lens needs to satisfy aesthetic and
pragmatic design constraints. In many situations, such as for the
construction of car lights, physical molds are built by milling and
the mirror or lens is built on this mold. Sometimes the optical
component itself is directly milled. This imposes some constraints
that can be achieved by imposing convexity or smoothness
conditions. The convexity constraint is classical since it allows in
particular to mill the component with a tool of arbitrary
large radius. Conversely, concavity allows to mill its mold.
Also, convex mirrors are easier to chrome-plate, because convex
surfaces have no bumps in which the chrome would spuriously
concentrate~\cite{cork1977method}.

\begin{figure}[t]
  \includegraphics[width=\linewidth]{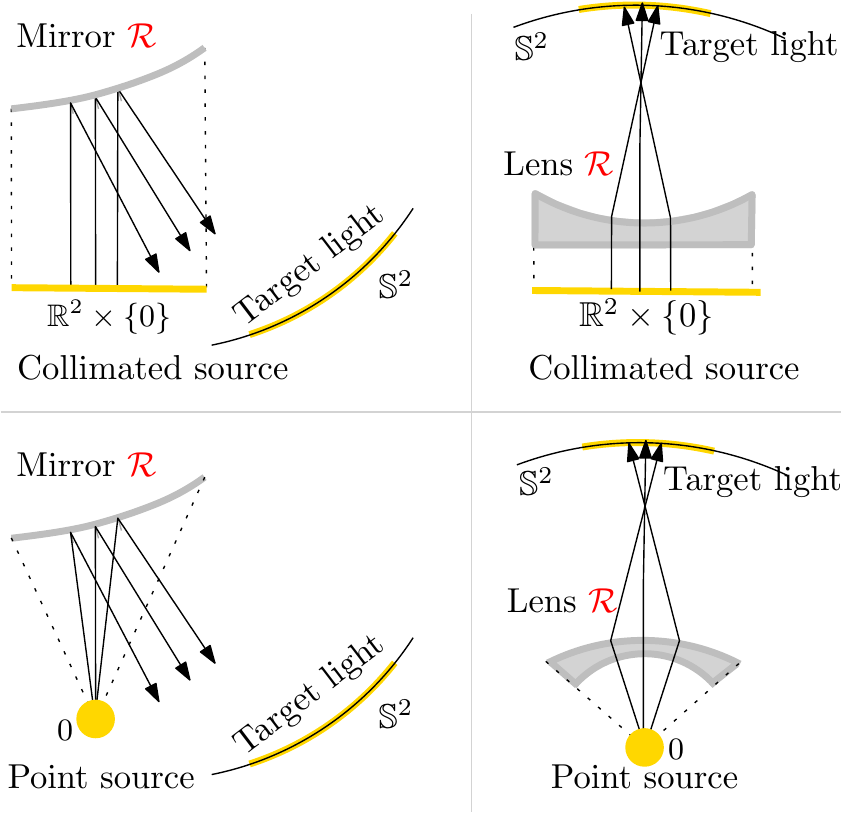}

  \caption{\textbf{Four inverse problems arising in non-imaging
      optics.} In each case, the goal is to build the surface $\Ref$
    of a mirror or a lens.  For each problem, we provide two solutions
    (for instance, convex or concave surfaces when the light source is
    collimated). \textit{Top/Bottom:} Collimated light sources/Point
    light sources. \textit{Left/Right:} Mirror/Lens
    design.\label{figure:examples-inverseproblems}}
  
\end{figure}

In this paper, we propose a generic algorithm capable of solving eight
different caustic design problems, see Figures~\ref{fig:teaser} and
\ref{figure:examples-inverseproblems}. Our approach relies on the
relation between these problems and optimal transport. The algorithm
is fully generic in the sense that it can deal with any of the eight
caustic-design problems just by changing a formula, and can handle
virtually any ideal light source and target.
Our contributions are the following:
\begin{itemize}
\item
We propose a general framework for eight different optical component
design problems (i.e. four non-imaging problems, for which we can
produce either concave or convex solutions). These problems amount to
solving the same  \jocelynbis{\textit{light energy
    conservation} equation (see Sec.~\ref{sec:LEC})}, which involves
prescribing the amount of light reflected or refracted in a finite
number of directions.
\item 
We propose a single algorithm with no parameter capable to solve
\jocelynbis{this equation for}
the eight different problems.
\jocelynbis{We will see that, in order to solve this equation, we need to
    compute}
integrals over \emph{visibility cells}, which can be obtained in all
cases by intersecting a 3D Power diagram with a planar of spherical
domain. The equation is then solved using a damped Newton algorithm.
\item In all of the four non-imaging problems, we can construct either
  a concave or convex optical component, easing their
  fabrication. Several components can then be combined to produce a
  single caustic, providing resilience to small obstacles and providing
  degrees of freedom to  to control the shape of the
  optical system.  \jocelyn{
\item We show that we can solve near-field problems (when the target is at a
    finite distance) by iteratively solving far-field problems (when the target
    is a set of directions) for the eight optical component design problems.
}
\end{itemize}

\section{Related work}
The field of non-imaging optics has been extensively studied in the last
thirty years. We give below an overview of the main approaches to
tackle several of the problems of this field.  A survey on
inverse surface design from light transport behaviour is provided by
\citeauthor{patow2005survey} \shortcite{patow2005survey}.

\paragraph{Energy minimization methods.}
Many different methods to solve inverse problems arising in non-imaging optics
rely on variational approaches. When the energies to be minimized are not
convex, they can be handled by different kind of iterative methods.
One class of methods deals with stochastic optimization.
\citeauthor{finckh2010geometry} \shortcite{finckh2010geometry} propose
to represent the optical component (mirror or lens) as a $\Class^2$
B-spline triangle mesh and to use stochastic optimization to adjust
the heights of the vertices so as to minimize a light energy
constraint.  Note that this approach is very costly, since a forward
simulation needs to be done at every step and the number of steps is
very high in practice.  Furthermore, using this method, lots of
artifacts in the final caustic images are present.
%
%
%
%
Stochastic optimization has also been used by \citeauthor{papas2011goal}
\shortcite{papas2011goal} to
design reflective or refractive caustics for collimated light
sources. At the center of the method is the Expectation Minimization
algorithm initialized with a Capacity Constrained Voronoi Tessellation
(CCVT) using a variant of  Lloyd's algorithm
\cite{lloyd1982least}. The source is a uniform directional light and
is modeled using an array of curved microfacets.  The target is
represented by a mixture of Gaussian kernel functions. This method
cannot accurately handle low intensity regions and artifacts due to
the discretization are present.  Microfacets were also used by
\citeauthor{weyrich2009fabricating} \shortcite{weyrich2009fabricating} to represent the mirror. Due to the
sampling procedure, this method cannot correctly handle smooth regions
and does not scale well with the size of the target.  
  \jocelynbis{More
  recently, \citeauthor{piovarvci2017directional}
  \shortcite{piovarvci2017directional} used
  microgeometry to design directional screens which provide increased
  gain and brightness. In their approach, the screen is decomposed into 
  many small patches, each patch reflecting a set of rays toward a prescribed cone of directions. 
 Their problem for each patch is similar to the one we solve for the special case of a directional
  source and a target at infinity and corresponds to one pillow, see Section~\ref{sec:results}. Similarly to us, 
  their approach is based on convex optimization and produce convex patches. They prescribe the areas 
  of the facets, whereas we prescribe their measures. Their numerical approach relies on gradient descent rather
  than on Newton's method, and only deals with collimated source and far-field target.}
  
The approaches proposed by \citeauthor{kiser2012architectural}
\shortcite{kiser2012architectural}, \citeauthor{yue2014poisson}
\shortcite{yue2014poisson} and \citeauthor{schwartzburg2014high}
\shortcite{schwartzburg2014high} have in common that they first compute some kind of relationship between
the incident rays and their position on the target screen and then use an
iterative method to compute the shape of the refractive surface.
\citeauthor{yue2014poisson} \shortcite{yue2014poisson} use a continuous parametrization and thus
cannot correctly handle totally black and high-contrast regions (boundaries between
very dark and very bright areas).
\citeauthor{yue2012pixel} \shortcite{yue2012pixel} proposed to use sticks to represent the
refractive surface. This allows to reduce production cost, to be more
entertaining for the user since a single set of sticks can produce different
caustic patterns. The main problem with this approach is the computational
complexity since they need to solve a NP-hard assignment problem.
The problem of designing lenses for collimated light sources has also
been considered by \citeauthor{schwartzburg2014high}
\shortcite{schwartzburg2014high}. They propose a method to build
lenses that can refract complicated and highly contrasted targets. They first use
optimal transport on the target space to compute a mapping between the refracted
rays of an initial lens and the desired normals, then perform a post-processing step 
to build a surface whose normals are close to the desired ones.

\paragraph{Monge-Amp\`ere equations} When the source and target lights are
modeled by continuous functions, the problem amounts to solving a
generalized Monge-Amp\`ere equation, either in the plane for
collimated light sources, or on the sphere for point light sources.
\jocelynbis{Let us explain this link more precisely for a collimated
  light source, assuming that the source rays are collinear to the
  constant vector $e_z=(0,0,1)$ and emitted from a horizontal domain $\Omega
  \subset \Rsp^2\times \{0\}$. We assume that the optical component is
  smooth and parameterized by a height-field function
  $\varphi:\Omega\to \Rsp$ and denote by $\mu$ and $\nu$ the source
  and target measures. At every point $\varphi(x)$ of the optical
  component, the gradient $\nabla \varphi(x)$ encodes the normal, and
  we denote by $F(\nabla \varphi(x)) \in \Sph^2$ the direction of the
  ray that is reflected at $(x, \phi(x))$ using Snell's law. The
  conservation of light energy thus reads $\nu(A)=\mu((F\circ \nabla
  \varphi)^{-1}(A))$ for every set $A$. This is equivalent to having
  $\tilde{\nu}(B)=\mu((\nabla \varphi)^{-1}(B))$ for every set $B$,
  where $\tilde{\nu}(B)=\nu(F(B))$. When $F$ and $\nabla \varphi$ are
  one-to-one (which is the case if the optical component is convex or
  concave) and $\mu$ and $\tilde{\nu}$ are modeled by continuous
  functions $f$ and $g$, with the change of variable formula, the
  light energy conservation becomes equivalent to the following
  generalized Monge-Amp\`ere equation
\begin{equation} \label{eq:cma}
g(\nabla \varphi (x)) \det ( D^2 \varphi (x)) = f(x).
\end{equation}
Similar equations are obtained for point light sources.}
The existence and regularity of their solutions, namely of the mirror
or lens surfaces, have been extensively studied. When the light source
is \jocelynbis{a point}, this problem has been studied for
mirrors~\cite{caffarelli2008weak,caffarelli2008regularity} and
lenses~\cite{gutierrez2009refractor} and when the light source is
collimated one recovers \eqref{eq:cma}~\cite{gutierrez2013parallel}.  We refer to the book of
\citeauthor{gutierrez2016monge} \shortcite{gutierrez2016monge} for an introduction to
Monge-Amp\`ere equations.

%

\paragraph{Optimal transport based methods in non-imaging optics} In fact, the
Monge-Amp\`ere equations corresponding to the non-imaging problems considered in
this paper can be recast as optimal transport problems. This was first observed
by \citeauthor{wang2004design} \shortcite{wang2004design} and
\citeauthor{glimm2003optical} \shortcite{glimm2003optical} for the mirror problem with a point light source. Many algorithms related to optimal transport have been developed to address  non-imaging problems. For collimated sources, one could rely on wide-stencils finite difference schemes~\cite{prins2013numerical},
or on numerical solvers for quadratic semi-discrete optimal transport
\cite{merigot2011multiscale,de2012blue}. For point
sources, there exist variants of the Oliker-Prussner algorithm for the
mirror problem~\cite{caffarelli1999numerical} or the lens
problem~\cite{gutierrez2009refractor}. Both algorithms have a
$\mathrm{O}(N^4)$ complexity, restricting their use to small
discretizations. A quasi-Newton method is proposed by
\citeauthor{de2015far} \shortcite{de2015far} for point-source reflector design, handling up to  $10^4$ Dirac masses.

Finally we note that the approach of \citeauthor{schwartzburg2014high}
\shortcite{schwartzburg2014high} to
build lenses also relies on optimal transport. However, the optimal
transport step is used as a heuristic to estimate the normals of the
surface, and not to directly construct a solution to the non-imaging
problem.  A post-processing step is then performed by minimizing a
non-convex energy composed of five weighted terms. In contrast, all
the results presented in this article use no post-processing.
\section{Light energy conservation}\label{sec:LEC}
We present in this section several mirror and lens design problems
arising in non-imaging optics. In all the problems, one is given a
light source (emitted by either a plane or a point) and a desired
illumination ``at infinity'' after reflection or refraction, which is
called the target, and the goal is to design the geometry of a mirror
or lens which transports the energy emitted by the source onto the
target.  \jocelyn{We do not take into account multiple reflections or
  refractions.}  We show that even though the problems we consider are
quite different from one another, they share a common structure that
corresponds to a so-called generalized Monge-Amp\`ere equation, whose
discrete version is given by Equation \eqref{eq:LEC}.
\jocelynbis{This section gathers and reformulates in a unified setting
  results about mirror and lens design for collimated and point light
  sources. We refer to the work of \citeauthor{caffarelli2008weak} \shortcite{caffarelli2008weak}, \citeauthor{prins2013numerical} \shortcite{prins2013numerical},
  \citeauthor{de2015far}  \shortcite{de2015far} and references therein.}


\subsection{\jocelynbis{Collimated light source}}

\subsubsection{\jocelynbis{Convex mirror design}}
\begin{figure}
  \centering
  \includegraphics[width=.8\linewidth]{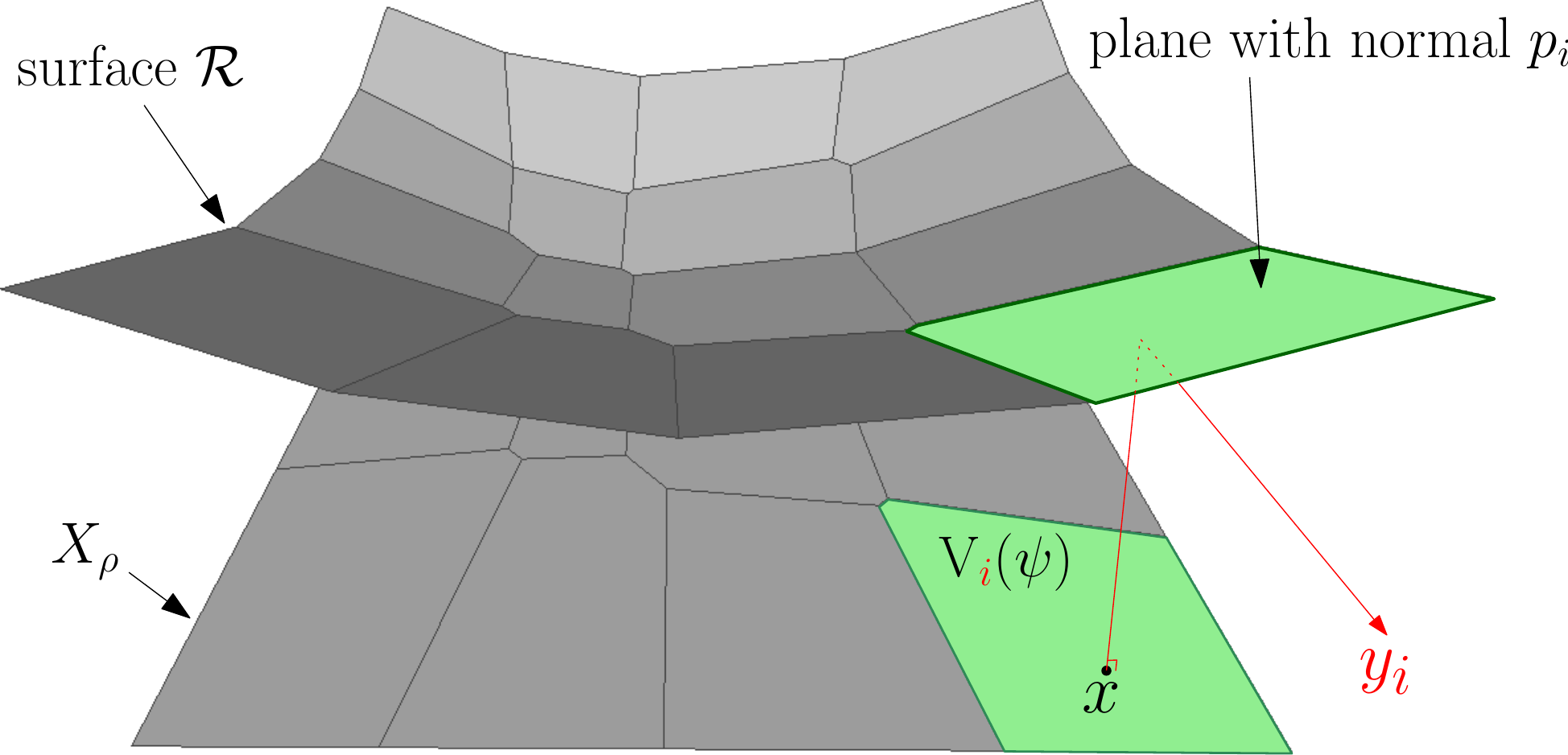}
  \caption{\textbf{Convex Mirror for a collimated light source} (when $N=16$). The mirror surface $\Ref$ is the graph of a convex piecewise affine
  function. The support $X_\rho$ of $\rho$ is decomposed into visibility cells
  $(V_i(\psi))_{1\leq i \leq N}$. Every vertical ray above a point $x\in X_\rho$
  belongs to a cell $V_i(\psi)$, touches a plane \jocelyn{with slope} $p_i$ and is reflected to the direction $y_i$.}\label{figure:collimated_reflector}
\end{figure}

In this first problem, the light source is collimated, meaning that it
emits parallel rays. The source is encoded by a light intensity
function $\rho:\Omega \to\Rsp$ over a 2D domain $\Omega$ contained in
the $(xy)$ plane $\Rsp^2\times \{0\} \subset \Rsp^3$, and that all the
rays are parallel and directed towards $e_z=(0,0,1)$. For simplicity,
we will conflate $\Rsp^2$ and $\Rsp^2\times\{0\}$. The desired target
illumination is ``at infinity'', and is described by a set of
intensity values $\sigma=(\sigma_i)_{1\leq i \leq N}$ supported on a
finite set of directions $Y=\{y_1,\cdots,y_N\}$ included in the unit
sphere $\Sph^2$.  The problem is to find the surface $\Ref$ of a
mirror that reflects the source intensity $\rho$ to the target
intensity $\sigma$, see Figures~\ref{figure:examples-inverseproblems}
(top left) and \ref{figure:collimated_reflector}. 

Since the number of reflected directions $Y$ is finite, the mirror
surface $\Ref$ is composed of a finite number of planar facets as
illustrated in Figure~\ref{figure:collimated_reflector}. We will
construct the mirror $\Ref$ as the graph of a piecewise-linear convex
function of the form $x \in\Rsp^2 \mapsto \max_i
\sca{x}{p_i} - \psi_i$, where $ \sca{\cdot}{\cdot} $ denotes the dot
product. The vectors $p_1,\hdots,p_N \in\Rsp^2$ and the elevations
$\psi = (\psi_1,\hdots,\psi_N)\in\Rsp^N$ have to be determined. To
choose $p_i$, we require that the plane $P_i = \{ (x, \sca{x}{p_i})
\mid x\in\Rsp^2\}$ reflects rays with direction $e_z$ towards the
direction $y_i\in\Sph^2$.  The downward pointing unit normal to $P_i$
is given by the formula $n_i = (p_i,-1)/(\nr{p_i}^2 +1) \in \Rsp^3$,
and Snell's law of reflection gives us $ y_i = e_z - 2 \sca{n_i}{e_z}
n_i$.  Solving for $p_i$ and denoting $\proj_{\Rsp^2}$ the orthogonal
projection onto $\Rsp^2 \times \{0\} $, we get
$$p_i= -\proj_{\Rsp^2}(y_i-e_z)/\sca{y_i-e_z}{e_z}.$$
Given a vector of elevations $\psi:=(\psi_i)_{1\leq i \leq N}$, we
define the visibility cell $\V_i(\psi)$
$$
\V_i(\psi)=\{x\in \Rsp^2\times\{0\} \mid~ \forall j, -\sca{x}{p_i} + \psi_i \leq -\sca{x}{p_j} + \psi_j \}.
$$ By construction, for every $i\in\{1,\hdots,N\}$, any vertical ray emanating from a point $x \in
V_i(\psi)$ touches the mirror surface $\Ref$ at an altitude
$\sca{x}{p_i} - \psi_i$, and  is thus reflected towards
direction $y_i$. Consequently, the amount of light reflected
towards direction $y_i$ equals the integral of $\rho$ over
$V_i(\psi)$.
The \textit{Collimated Source Mirror problem} \csmirror
then amounts to finding elevations $\psi \in\Rsp^N$ such that
\begin{equation}\label{eq:LEC}
    \forall i \in \{1,\cdots,n\}\quad  \int_{\V_i(\psi)}\rho(x)\d x=\sigma_i. 
\end{equation}
By construction, a solution to Equation (\ref{eq:LEC}) provides a
parameterization $\Ref_\psi$ of a convex mirror that reflects the
collimated light source $\rho$ to the discrete target $\sigma$:
$$
\Ref_{\psi}:\ x\in \Rsp^2\ \mapsto\ (x,\max_i \sca{x}{p_i} - \psi_i) \in\Rsp^3,
$$ where $ \Rsp^2 \times \{0\} $ and $\Rsp^2$ are identified. Notice
that since the mirror is a graph over $\Rsp^2\times\{0\}$, the vectors
$y_i$ cannot be upward vertical. In practice we assume that every
direction $y_i$ belongs to the hemisphere $\Sph^2_{-}:=\{x\in
\Sph^2,\ \sca{x}{e_z} \leq 0\}$. Furthermore, we localize the position
of the mirror by considering it only above the domain $X_\rho:=\{x\in
\Rsp^2\times \{0\},~\rho(x)\neq 0\}$.

\paragraph{Concave mirror.} The same approach also allows the construction of concave mirrors,
using a concave function of the form $ x\mapsto \min_i \sca{x}{p_i} +
\psi_i$. This amounts to replacing the visibility cells by
$$ \V_i(\psi)=\{x\in \Rsp^2\times\{0\} \mid~ \sca{x}{p_i} + \psi_i
\leq \sca{x}{p_j} + \psi_j~ \forall j\}.
$$ In that case, a solution to Equation (\ref{eq:LEC}) provides a
parametrization of a concave mirror $ \Ref_{\psi}(x)=(x,\min_i
\sca{x}{p_i} + \psi_i)$ that sends the collimated light source $\rho$
to the discrete target $\sigma$.

\subsubsection{\jocelynbis{Convex lens design}}
\jocelynbis{In this second design problem, we are interested in designing lenses that refract a given
collimated light source intensity $ \rho $ to a target intensity $\sigma$,
see the top right diagram in Figure \ref{figure:examples-inverseproblems}.
We denote by $n_1$ the refractive index of the lens, by $n_2$ the ambient space refractive
index and by $\kappa=n_1/n_2$ the ratio of the two indices.}
We assume that the rays
emitted by the source are vertical and that the bottom of the lens is
flat and orthogonal to the vertical axis. There is no refraction
angle when the rays enter the lens, and we  only need to
build the top part of the lens.

By a simple change of variable, we show that this problem is
equivalent to \csmirror. More precisely, for every $y_i\in Y$, we now
define $p_i$ to be the \jocelyn{slope} of a plane that refracts the vertical ray
$e_z$ to the direction $y_i$.
We define $\Ref$ as the graph of a convex
function of the form $x\mapsto\max_i \sca{x}{p_i} - \psi_i$, where
$\psi=(\psi_i)_{1\leq i \leq N}$ is the set of elevations.
\jocelynbis{A calculation similar to the \csmirror case gives the following
    expression:
    $$p_i= -\proj_{\Rsp^2}(y_i- \kappa e_z)/\sca{y_i- \kappa e_z}{e_z}.$$
In that case,} we define the visibility cell $\V_i(\psi)$ to be the set of points $x\in \Rsp^2
\times \{0\} $ that are refracted to the direction $y_i$:
$$
\V_i(\psi)=\{x\in \Rsp^2\times \{0\} \mid \forall j,~ -\sca{x}{p_i} + \psi_i \leq -\sca{x}{p_j} + \psi_j\}.
$$
The \textit{Collimated Source Lens problem} \cslens then amounts to finding weights
$(\psi_i)_{1\leq i \leq N}$ that satisfy (\ref{eq:LEC}).  In that case the lens surface is parameterized by
$$\Ref_\psi:\ x\in \Rsp^2\ \mapsto\ (x,\max_i \sca{x}{p_i} - \psi_i).$$
In practice, we choose the directions $y_i$ in $\Sph^2_{+}$ and the mirror to be parameterized over the support $X_\rho$ of $\rho$.

\begin{figure}
    \centering
    \includegraphics[width=.48\linewidth]{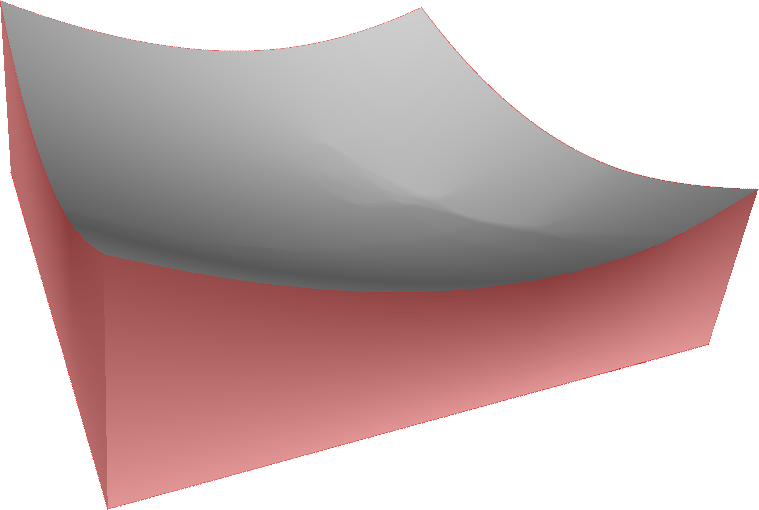}
    \includegraphics[width=.48\linewidth]{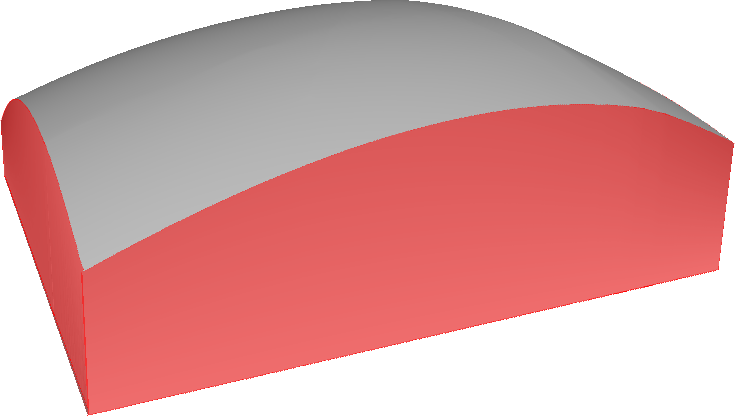}
    \caption{
          \label{fig:res-cs_lens-concave-convex}\textbf{Concave and convex
              lenses} for a uniform collimated light source and the same
          target.}
\end{figure}

\paragraph{Concave lens.}
\jocelynbis{Note} that we can also build \jocelyn{concave} lenses by considering
parameterizations with \jocelynbis{concave} functions of the form $x\mapsto\min_i
\sca{x}{p_i} + \psi_i$. Figure~\ref{fig:res-cs_lens-concave-convex}
illustrates a concave and a convex solution to the same non-imaging
optics problem.

\subsection{\jocelynbis{Point light source}}

\subsubsection{\jocelynbis{Concave mirror design.}}
In this second mirror design problem, all the rays are now emitted from a
single point in space, located at the origin, and the light source is
described by an intensity function $\rho$ on the unit sphere
$\Sph^2$. As in the previous cases, the target is ``at infinity'' and
is described by a set of intensity values $\sigma=(\sigma_i)_{1\leq i
  \leq N}$ supported on the finite set of directions $Y=\{y_1,\cdots,
y_N\}\subset \Sph^2$. The problem we consider is to find the surface
$\Ref$ of a mirror that sends the light intensity $\rho$ to the light
intensity $\sigma$ (Fig.~\ref{figure:examples-inverseproblems},
bottom left).

Following~\citeauthor{caffarelli2008weak} \shortcite{caffarelli2008weak}, we build a concave surface $\Ref$
that is composed of pieces of confocal paraboloids. More precisely, we
denote by $P(y_i,\psi_i)$ the solid (i.e filled) paraboloid whose
focal point is at the origin with focal distance $\psi_i$ and with
direction $y_i$. We define the surface $\Ref_\psi$ as the boundary of
the intersection of the solid paraboloids, namely $\Ref_\psi=\partial
\left(\cap_i P(y_i,\psi_i)\right)$. The visibility cell $\V_i(\psi)$
is the set of ray directions $x\in \Sph^2$ emanating from the light
source that are reflected in the direction $y_i$. Since each
paraboloid $\partial P(y_i,\psi_i)$ is parameterized over the sphere
by $x\mapsto \psi_i x/ (1-\sca{x}{y_i})$, one
has~\cite{caffarelli2008weak}
$$
\V_i(\psi)=\left\{x\in \Sph^2 \mid~ \forall j,  \frac{\psi_i}{1-\sca{x}{y_i}} \leq \frac{\psi_j}{1-\sca{x}{y_j}}  \right\}.
$$ The \textit{Point Source Mirror problem} \psmirror then amounts to
finding  $(\psi_i)$ that satisfy the light energy conservation
equation (\ref{eq:LEC}). The mirror surface is then parameterized by
$$\Ref_\psi:\ x\in \Sph^2\ \mapsto\ \min_i
\frac{\psi_i}{1-\sca{x}{y_i}}\ x.$$ In practice, we assume that the
target $Y$ is included in $\Sph^2_{-}$, that the support $X_\rho$ of
$\rho$ is included $\Sph^2_+:=\{x\in \Sph^2,\ \sca{x}{e_z}\geq 0\}$,
and that the mirror is parameterized over $X_\rho$.

One can also define the mirror surface as the boundary of the union (instead of the intersection) of a family of solid paraboloids. Then, the visibility cells become
$$
\V_i(\psi)=\left\{x\in \Sph^2 \mid~ \forall j, \frac{\psi_i}{1-\sca{x}{y_i}} \geq \frac{\psi_j}{1-\sca{x}{y_j}} \right\}
$$ and a solution to Equation~(\ref{eq:LEC}) provides a
parameterization $\Ref_\psi(x) = x \max_i \psi_i/(1 - \sca{x}{y_i})$
of the mirror surface. \jocelynbis{Let us note that in this case the mirror is
    neither convex nor concave.}

\subsubsection{\jocelynbis{Convex lens design.}}
We now consider the \jocelynbis{lens design problem for a point light source.}
As in the collimated
setting, we fix the bottom part of the lens. We choose a piece of
sphere centered at the source, so that the rays are not deviated. Following
~\citeauthor{gutierrez2009refractor} \shortcite{gutierrez2009refractor}, the lens is composed of pieces of
ellipsoids of constant eccentricities $ \kappa > 1 $, where $\kappa $ is
the ratio of the indices of refraction. 
Each ellipsoid $\partial E(y_i,\psi_i)$ can be parameterized over the
sphere by $x\mapsto \psi_i x/ (1-\kappa \sca{x}{y_i})$. The visibility
cell of $y_i$ is then
$$
\V_i(\psi)=\left\{x\in \Sph^2 \mid~ \forall j, \frac{\psi_i}{1-\kappa\sca{x}{y_i}} \leq \frac{\psi_j}{1-\kappa \sca{x}{y_j}} \right\}.
$$ The \textit{Point Source Lens problem} \pslens then amounts to
finding weights $(\psi_i)_{1\leq i \leq N}$ that satisfy
(\ref{eq:LEC}). \jocelynbis{Note} that the top surface of the lens is then
parameterized by
$$\Ref_\psi:\ x\in \Sph^2\ \mapsto\ \min_i \frac{\psi_i}{1 - \kappa \sca{x}{y_i}}\ x.$$
In practice, we choose the set of directions $y_i$ to belong to $\Sph^2_{+}$ and the lens to be parameterized over the support $X_\rho\subset \Sph^2_+$ of $\rho$.

One can also choose to define the lens surface as the boundary of the union (instead of the intersection) of a family of solid ellipsoids. In that case, the visibility cells are given by
$$
\V_i(\psi)=\left\{x\in \Sph^2 \mid~ \forall j, \frac{\psi_i}{1-\kappa\sca{x}{y_i}} \geq \frac{\psi_j}{1-\kappa\sca{x}{y_j}} \right\}
$$
and a solution to Equation~(\ref{eq:LEC}) provides a parameterization $\Ref_\psi(x) = x \max_i \psi_i/(1 - \kappa \sca{x}{y_i})$ of the lens surface.
\jocelynbis{Let us note that in this case the lens is neither convex nor
    concave.}

\subsection{General formulation}

Let $X$ be a domain of either the plane $\Rsp^2\times\{0\}$ or the
unit sphere $\Sph^2$, $\rho:X\to\Rsp$ a probability density and
$Y=\{y_1,\cdots,y_N\}\subset \Sph^2$ be a set of $N$ points. We define
the function $G:\Rsp^N\to \Rsp^N$ by
$$
G_i(\psi)=\int_{\V_i(\psi)}\rho(x) \d x
$$ where $G(\psi) = (G_i(\psi))_{1\leq i \leq N}$ and $\V_i(\psi)
\subset X$ is the visibility cell of $y_i$, whose definition depends
on the non-imaging problem.  Using this notation, Equation
\eqref{eq:LEC} can be rephrased as finding weights
$\psi=(\psi_i)_{1\leq i \leq N}$ such that
\begin{equation}\label{eq:generalProblem}
\forall i \in \{1,\cdots,N\},\quad G_i(\psi) = \sigma_i.
\end{equation}


Many other problems arising in non-imaging optics amount to solving
equations of this form. For example, the design of a lens that
refracts a point light source to a desired near-field target can also
be modeled by a Monge-Amp\`ere equation that has the same
structure~\cite{gutierrez2009refractor}.  In this case, the visibility
diagram correspond to the radial projection onto the sphere of pieces
of confocal ellipsoids with non constant eccentricities and is not
associated to an optimal transport problem.

\section{Visibility and Power cells}\label{sec:visibility-cells-power-diagram}

The main difficulty to evaluate the function $G$ appearing in
Equation~\eqref{eq:generalProblem} is to compute the visibility cells
$V_i(\psi)$ associated to each optical modeling problem.  We show in
this section that the visibility cells have always the same structure,
allowing us to build a generic algorithm in Section~\ref{sec:algo}.
\jocelynbis{We first need to introduce the notion of \emph{Power
    diagram}.}

\jocelynbis{\paragraph{Power diagrams.} Let $P \subseteq
  \Rsp^3\times\Rsp$ be a weighted point cloud, i.e. $P = \{ (p_i,
  \omega_i) \}_{1\leq i\leq N}$ with $p_i\in\Rsp^2$ and
  $\omega_i\in\Rsp$. The \emph{Power cell} of the $i$th point $ p_i $
  is given by
$$ \Pow_i(\P) := \{ x \in \Rsp^3 \mid \forall j,~ \nr{x-p_i}^2 +
  \omega_i \leq \nr{x-p_j}^2 + \omega_j \}.  $$ Power cells partition
  $ \Rsp^3 $ into convex polyhedra up to a negligible set.  Power
  diagrams are well-studied objects appearing in computational
  geometry~\cite{aurenhammer1987power}, and can be computed
  efficiently in dimension $2$ and $3$.  When all the weights are equal, one
  recover the usual \emph{Voronoi diagram}.}

\paragraph{Visibility diagram as a restricted Power diagram.}
\jocelynbis{We now show that}
in all the non-imaging problems of
Section~\ref{sec:LEC}, the visibility cells are of the form
\begin{equation}\label{eq:pow}
V_i(\psi)=\Pow_i(\P) \cap X.
\end{equation}
For a collimated source, $X$ denotes the plane $\Rsp^2\times \{0\}$
and for a point source, $X$ is the unit sphere $\Sph^2$.
The expression of the weighted point cloud $\P = \{ (p_i,\omega_i) \}$
depends on the problem.  We refer to Table~\ref{tbl:formulas} and
the work of \citeauthor{de2015far} \shortcite{de2015far} for formulas in the \jocelynbis{\psmirror case, the
  other ones being obtained in a similar fashion.} Let us show the
derivation of the formula in \jocelynbis{the \csmirror case,} where
the $i$th visibility cell is given by
$$
\begin{aligned}
  \V_i(\psi)&=
  \{x\in \Rsp^2\times\{0\} \mid~ \forall j, -\sca{x}{p_i} + \psi_i \leq -\sca{x}{p_j} + \psi_j \} \\
  &=
\{x\in \Rsp^2\times\{0\} \mid~ \forall j, \nr{x-p_i}^2 + \omega_i \leq \nr{x-p_j}^2 + \omega_j \}, 
\end{aligned}
$$ where $\omega_i = 2\psi_i - \nr{p_i}^2$.  We conclude that the
visibility cells for a convex mirror of the \csmirror problem are indeed given by
\eqref{eq:pow}, where the weighted point cloud is given by the first
line of Table~\ref{tbl:formulas}.


\begin{table}
      \caption{Formulas for the weighted points used to define the Power
    cells in Equation \eqref{eq:pow} for the various problems. In the
    lens design problem, $\kappa > 0 $ is the ratio of the indices of
    refraction, $\kappa>1$ in the \pslens setting. Ccv means concave
    and Cvx convex.  $\widetilde{\mathrm{Ccv}}$ means that the optical
    component converges to a concave one when the discretization tends
    to infinity. \label{tbl:formulas}}
  \resizebox{\linewidth}{!}{
    \begin{tabular}{|c|l|l|}
  \hline
  \textbf{Type} & Points & Weights \\
  \hline
  Cvx \csmirror & $\jocelynbis{p_i=-\frac{\proj_{\Rsp^2}(y_i-e_z)}{\sca{y_i-e_z}{e_z}}}$ &
  $\omega_i = 2\psi_i - \nr{p_i}^2.$ \\
  Ccv \csmirror & $\jocelynbis{p_i= \frac{\proj_{\Rsp^2}(y_i-e_z)}{\sca{y_i-e_z}{e_z}}}$ &
  $\omega_i = 2\psi_i - \nr{p_i}^2.$ \\
  Cvx \psmirror & $p_i = -\frac{y_i}{2\ln(\psi_i)}$ &  $\omega_i = -\frac{1}{\ln(\psi_i)} -
\frac{1}{4\ln(\psi_i)^{2}}$ \\
  $\widetilde{\mathrm{Ccv}}$ \psmirror & $p_i = y_i/(2\ln(\psi_i))$ & $\omega_i = \frac{1}{\ln(\psi_i))} - \frac{1}{4\ln(\psi_i)^{2}}$ \\
  Cvx \cslens & $p_i= -\frac{\proj_{\Rsp^2}(y_i-\kappa e_z)}{\sca{y_i-\kappa e_z}{e_z}}$
  &  $ \omega_i = 2\psi_i - \nr{p_i}^2$ \\
  Ccv \cslens & $p_i= \frac{\proj_{\Rsp^2}(y_i-\kappa e_z)}{\sca{y_i-\kappa e_z}{e_z}}$
  &  $ \omega_i = 2\psi_i - \nr{p_i}^2$ \\
  Cvx \pslens & $p_i = -\kappa\frac{y_i}{2\ln(\psi_i)}$ &
  $\omega_i =
  -\frac{1}{\ln(\psi_i)} - \frac{\kappa^2}{4\ln(\psi_i)^{2}}$\\
  $\widetilde{\mathrm{Ccv}}$ \pslens & $p_i = \kappa y_i / (2\ln(\psi_i))$ & $\omega_i = \frac{1}{\ln(\psi_i)} - \frac{\kappa^2}{4\ln(\psi_i)^{2}}$\\
  \hline
    \end{tabular}
  }
\end{table}


\section{A generic algorithm}\label{sec:algo}
For each optical design problem, given a light source intensity
function, a target light intensity function and a tolerance,
Algorithm~\ref{algo:main} outputs a triangulation of a mirror or a
lens that satisfies the light energy conservation equation \eqref{eq:LEC}.

The main problem is to find weights $\psi$ such that $G(\psi) =
\sigma$ (see Equation \eqref{eq:generalProblem}). This is done using a
damped Newton algorithm similar to recent algorithms that have been
shown to have a quadratic local convergence rate for optimal transport
problems~\cite{kitagawa2016newton} or for Monge-Amp\`ere equations in
the plane~\cite{mirebeau2015discretization}.  A key point of
  this algorithm is to enforce the Jacobian matrix $ \D G(\psi) $ to
  always be of rank $N-1$. To this purpose, we need to enforce all
along the process that
 \begin{equation}
   \label{eq:damping}
   \forall i \in\{1,\hdots,N\},~G_i(\psi)>0.
 \end{equation}
Indeed, first note that since $ G $ is invariant under the addition of a constant,
 the kernel of $ \D G(\psi) $ always contains the constant vector $ (1,\ldots, 1)$.
Now note that if we have $G_i(\psi) = 0$, then the corresponding visibility cell
$V_i(\psi)$ is empty, which implies that $ \nabla G_i(\psi) =
0$ (the gradient being taken with respect to $ \psi $). This is because the gradient of $G_i$ involves integral on the
boundary $\partial V_i(\psi)$, as shown for instance by
\citeauthor{kitagawa2016newton} \shortcite{kitagawa2016newton} in Theorem~1.3.
Hence, if $G_i(\psi) = 0$, then the
rank of $ \D G(\psi) $ is at most $N-2$ which prevents from using the Damped
Newton method.
Our method consists of three steps, described in Algorithm~\ref{algo:main}:

\begin{itemize}
\item \textbf{Initialization} (Sec. \ref{subsec:initialization}): We
  first discretize the source density into a piecewise affine density
  and the target one into a finitely supported measure. Then, we
  construct initial weights $\psi^{0}$ satisfying 
  $\forall i, G_i(\psi^{(0)}) > 0.$
\item \textbf{Damped Newton} (Sec \ref{subsec:newton}): We construct a
  sequence $\psi^{k}$ following Algorithm~\ref{algo:newton} until
  $\nr{G(\psi^{k}) - \sigma}_\infty \leq \eta$. The main difficulty
  here is to evaluate $G(\psi^{k})$ and $\DD G(\psi^{k})$.
\item \textbf{Surface construction} (Sec
  \ref{sec:surface-construction}): Finally, we convert the solution
  $\psi^{k}\in \Rsp^N$ into a triangulation. Depending on the
  non-imaging problem, this amounts to approximating an intersection
  (or union) of half-spaces (or solid paraboloids, or ellipsoids) by a
  triangulation.
\end{itemize}

\begin{algorithm}[t]
\begin{description}
  \item[Input] A light source intensity function $\rho_{in}$.\\
\hspace{.4cm} A target light intensity function $\sigma_{in}$.\\
\hspace{.4cm} A tolerance $ \eta > 0 $. \\
\item[Output] A triangulation $\Ref_T$ of a mirror or lens.
\item[Step 1] Initialization \hfill(Section \ref{subsec:initialization})\\
$ T, \rho \leftarrow $ \texttt{DISCRETIZATION\_SOURCE($\rho_{in})$}  \\
$ Y, \sigma \leftarrow $ \texttt{DISCRETIZATION\_TARGET($\sigma_{in}$)}  \\
$ \psi^0 \leftarrow $ \texttt{INITIAL\_WEIGHTS($ Y $)} \\
\item[Step 2] Solve Equation (\ref{eq:generalProblem}): $G(\psi)=\sigma$  \hfill(Section \ref{subsec:newton})\\
$\psi  \leftarrow$  \texttt{DAMPED\_NEWTON($T, \rho, Y, \sigma, \psi^0, \eta$)} \\
\item[Step 3] Construct a triangulation $\Ref_T$ of $\Ref$ \hfill(Section
\ref{sec:surface-construction})\\
$ \Ref_T \leftarrow $ \texttt{SURFACE\_CONSTRUCTION($\psi, \Ref_\psi$)} 
\end{description}
\caption{Mirror / lens construction \label{algo:main}}
\end{algorithm}

\subsection{Initialization}\label{subsec:initialization}
\paragraph{Discretization of light intensity functions}
Our framework allows to handle any kind of collimated or point light
source or target light intensity functions. It can be for example any
positive function on the plane or the sphere (depending on the
problem) or a greyscale image, which we see as piecewise affine
function. We first approach the support of the source density $\rho$
by a triangulation $T$ and assume that the density $\rho:T\to\Rsp^+$
is affine on each triangle. We then normalize $\rho$ by dividing it by
the total integral $\int_T \rho(x)\d x$.

Similarly, the target light intensity function can also be any
discrete probability measure. If the user provides an image, one can
transform it into a discrete measure of the form $ \sigma = \sum_i
\sigma_i \delta_{y_i} $ using Lloyd's algorithm or more simply by
taking one Dirac mass per pixel. We do the latter in all
experiments. The target measure is also normalized by dividing with
the discrete integral $\sum_i \sigma_i$. We need $\min_i \sigma_i > 0$
for the damped Newton algorithm, but this is not a
restriction: if $\sigma_i=0$, we simply remove the corresponding Dirac
mass $\delta_{y_i}$, thus ensuring that no light is sent towards
$y_i$.

\begin{algorithm}[t]
\begin{description}
  \item[Input] The source $\rho:T\to \Rsp^+$ and target $\sigma =
    \sum_i\sigma_i\delta_{y_i}$;
an initial vector $ \psi^0 $ and a tolerance $ \eta > 0 $. 
\item[Step 1] Transformation to an Optimal Transport problem\\
\begin{description}
\item[If $X=\Rsp^2\times \{0\}$,] then $\widetilde{\psi}^0=\psi^0$ (and $\widetilde{G}=G$).
\item[If $X=\Sph^2$,] then $\widetilde{\psi}^0=(\ln(\psi^0_i))_{1\leq i \leq N}$ (and $\widetilde{G}=(G_i\circ \exp)_{1\leq i \leq N}$).
\end{description}
\item[Step 2] Solve the  equation: $\widetilde{G}(\widetilde{\psi})=\sigma$\\
\begin{description}
\item[Initialization:]
 $\eps_0 := \min\left[\min_i G_i(\psi^0),~ \min_i \sigma_i\right] >  0$, 
  $k:=0$.
\item[While] $\nr{\widetilde{G}(\widetilde{\psi}^k) - \sigma}_\infty  \geq \eta$
  \begin{description}
\item[-] Compute $d_{k} = - \D \widetilde{G}(\widetilde{\psi}^k)^{+} (\widetilde{G}(\widetilde{\psi}^k) - \sigma)$
\item[-] Find the smallest $\ell \in \Nsp$ s.t. $\widetilde{\psi}^{k,\ell} :=
  \widetilde{\psi}^k + 2^{-\ell} d_k$ satisfies
\begin{equation*}
\left\{
\begin{aligned}
&\min_i \widetilde{G}_i(\widetilde{\psi}^{k,\ell})) \geq \eps_0 \\
&\nr{\widetilde{G}(\psi^{k,\ell}) - \sigma}_\infty \leq (1-2^{-(\ell+1)}) \nr{\widetilde{G}(\widetilde{\psi}^k) - \sigma}_\infty
\end{aligned}
\right.
\end{equation*}
\item[-] Set $\widetilde{\psi}^{k+1} = \widetilde{\psi}^k + 2^{-\ell}  d_k$ and $k\gets k+1$.
  \end{description}
\end{description}
\item[Return] $\psi:=(\widetilde{\psi}^{k}_i)_{1\leq i \leq N}$ if $ X =
    \Rsp^2 \times \{0\} $ or\\
    \hspace{.6cm} $\psi:=(\exp(\widetilde{\psi}^{k}_i))_{1\leq i \leq N}$ if $ X = \Sph^2 $.
\end{description}
\caption{Damped Newton method for $G(\psi)=\sigma$}
\label{algo:newton}
\end{algorithm}

\paragraph{Choice of the initial family of weights $\psi^0$.}
As mentioned at the beginning of this section, we need to ensure that
at each iteration all the visibility cells have non-empty
interiors. In particular, we need to choose a set of initial weights $
\psi^0 =(\psi^0_i)_{1\leq i \leq N}$ such that the initial visibility
cells are not empty.
\begin{itemize}
    \item For the collimated light sources cases \csmirror and
      \cslens, we see that if we choose $\psi^0_i = \nr{p_i}^2 / 2 $
      then $ \omega_i = 0 $, where $p_i$ is obtained using the
      formulas of the Section
      \ref{sec:visibility-cells-power-diagram}. Then, the visibility
      diagram becomes a Voronoi diagram, hence $ p_i \in \V_i(\psi^0)
      $.
    \item For the Point Source Mirror \psmirror case, an easy calculation shows that if we choose  $
        \psi^0_i = 1$, then $ -y_i \in \V_i(\psi^0) $.
    \item For the Point Source Lens \pslens case, we can show that if we also choose
$\psi^0_i = 1$, then $ y_i \in \V_i(\psi^0) $.
\end{itemize}
\jocelynbis{Note} that the previous expressions for $\psi^0$ ensure that
$G_i(\psi^0) = \rho(V_i(\psi^0)) > 0$ only when the support $X_\rho$
of the light source is large enough. As an example in the \psmirror
case, if $-y_i\notin X_\rho$, then we may have $G_i(\psi^0)=0$. To
handle this difficulty, we use a linear interpolation between $\rho$
and a constant density supported on a set that contains the
$(-y_i)$'s.  This strategy also works for the \csmirror, \pslens and
\cslens cases.

\subsection{Damped Newton algorithm}\label{subsec:newton}
When the light source is collimated (\textit{i.e.}
$X=\Rsp^2\times\{0\}$), the problem is known to be an optimal
transport problem in the plane for the quadratic cost, the function
$G$ is the gradient of a concave function, its Jacobian matrix $\D G$ is
symmetric and $\D G \leq 0$. Moreover, if $G_i(\psi)>0$ for all $i$
and if $X_\rho$ is connected, then the kernel of $\D G$ is spanned by
$\psi = \mathrm{cst}$. This ensures the convergence of the damped
Newton algorithm ~\cite{kitagawa2016newton} presented as
Algorithm~\ref{algo:newton}, where $A^+$ denotes the
\emph{pseudo-inverse} of the matrix $A$. Practically, taking the
pseudo-inverse of $\D \widetilde{G}(\widetilde{\psi}^k)$ guarantees
that the mean of the $\widetilde{\psi}^k$ remains constant. In
practice, we remove a line and a column of the matrix to make it full
rank.

When the light source is a point source, we make the change of
variable $\widetilde{\psi}=\jocelynbis{\ln}(\psi)$ and $\widetilde{G} = G\circ
\exp$, so that $G(\psi) = \sigma$ if and only if
$\widetilde{G}(\tilde\psi) = \sigma$. This change of variable turns
the optical component design problem into an optimal transport
problem, ensuring that $\widetilde{G}$ is the gradient of a concave
function and that $\D \widetilde{G}$ is symmetric negative
\cite{de2015far}, thus easily invertible. In the \psmirror problem
with convex mirrors, the damped Newton algorithm is also provably
converging~\cite{kitagawa2016newton}.

\begin{algorithm}[t]
\begin{description}
  \item[Input] The source $\rho:T\to \Rsp^+$ and target $\sigma =
    \sum_{i=1}^N \sigma_i\delta_{z_i}$; an initial vector $ \psi^0 $ and two tolerances $ \eta, \eta_{NF} > 0 $. \\
\item[Initialization] $ \forall i, c^0_i = O $
\item[While] $ \nr{c^{k+1} - c^{k}}_1/N > \eta_{NF} $
    \begin{description}
        \item[-] Compute $ v^{k}_i = \Ref_{\psi^{k}}(c^{k}_i) $
        \item[-] Set $ y^{k}_i = (z_i - v^{k}_i) / \nr{z_i - v^{k}_i} $
        \item[-] Solve $ \psi^{k+1} \leftarrow $ \texttt{SOLVE\_FF($T,\rho, Y^{k},
                \sigma, \eta$)}
        \item[-] Update $ c^{k+1}_i $ to be the centroid of $ V_i(\psi^{k+1}) $
    \end{description}
\end{description}
\caption{Optical component design for a NF target.\label{algo:near-field}}
\end{algorithm}

\paragraph{Computation of $G$ and $\D G$} By Section~\ref{sec:visibility-cells-power-diagram},
the visibility cells $V_i(\psi)$ can be computed by intersecting a
certain 3D Power diagram with a triangulation $T$ of the support
$X_\rho$ of $\rho$. Such intersection can be computed 
using the algorithm developed by \citeauthor{levy2014numerical}
\shortcite{levy2014numerical}. Then,
$$ G_i(\psi) = \int_{V_i(\psi)} \rho(x) \dd x $$ can be computed using first
order quadrature formulas. The computation of $\D G$ is done using
forward-mode automatic differentiation, where we store the gradient of
$G_i(\psi)$ as a sparse vector. Note that this works quite efficiently
since all numbers that occur in the computation of $G_i(\psi)$
depend only on the values $\psi_j$ where $j$ is such that $(i,j)$ are
neighbors in the visibility diagram, i.e.  $V_i(\psi) \cap V_j(\psi)
\neq \emptyset$.

\paragraph{Linear system}
Since $\D \widetilde{G}$ is sparse and symmetric negative, we solve
the linear systems  using preconditioned
conjugate gradient.


\subsection{Surface construction}
\label{sec:surface-construction}

In the last step of Algorithm~\ref{algo:main}, we build a
triangulation of the mirror or lens surface. The input is a family of weights $
\psi $ solving Equation \eqref{eq:generalProblem} and the
parameterization function $\Ref_\psi$ whose formula is given in
Section~\ref{sec:LEC} and depends on the eight different cases. We
triangulate each visibility cell by taking the convex hull of the vertices of
its boundary. A vertex of the triangulation
will belong to at least one visibility cell.
For each vertex, we can compute exactly
the normal to the (continuous surface) using Snell's law since we know
the incident ray and the corresponding reflected/refracted direction $
y_i $.


\section{Finite-distance caustics}
\label{sec:near-field}

In this section, we show that we can solve the eight optical component design problems when the target is 
at a finite distance. 
%
This setting is called \emph{near-field} (NF) in contrast with the previous \emph{far-field} (FF) case that deals with targets at infinity. 
This setting is interesting since in most applications, one wants the focused image to be at a finite distance and not at infinity.

To be more precise, given one of the eight optical component design problems
mentioned above with a target illumination $ \sigma = \sum_{i=1}^N \sigma_i
\delta_{z_i} $ supported on a set of points $ Z = \{ z_1, \ldots, z_N \} \subset
\Rsp^3 $, we propose an algorithm that iteratively solves a FF problem,
namely Equation~(\ref{eq:LEC}), and converges to a solution of the NF problem. 
%

\begin{table*}
  \begin{minipage}[t]{.49\linewidth}
    \begin{center}
      \caption{Running time and number of Newton steps in
        Algorithm~\ref{algo:newton} (FF target) for the \csmirror and
        the \textsc{Train} target in the far-field setting.      \label{tbl:ff-running-times} }
      \begin{tabular}{|c|c|c|}
        \hline
        \textbf{size} & \textbf{time} & \textbf{ \# Newton steps} \\
        \hline
        $128^2$ & 9s & 11 iterations \\
        $256^2$ & 38s & 13 iterations \\
        $512^2$ & 245s & 15 iterations \\
        $1024^2$ & 1598s & 18 iterations \\
        $2048^2$ & 7538s & 24 iterations \\
        \hline
      \end{tabular}
    \end{center}
  \end{minipage} \hfill
  \begin{minipage}[t]{.49\linewidth}
    \begin{center}
      \caption{Running time in Algorithm \ref{algo:near-field} (NF
        target) for the \csmirror and the \textsc{Train} target in the
        near-field setting for different
        discretizations.  \label{tbl:nf-running-times}}
      \begin{tabular}{|c|c|c|c|c|c|}
        \hline
        \textbf{size} & \textbf{$k=1$} & \textbf{$k= 2$} & \textbf{$k=3$} &
        \textbf{$k=4$} & \textbf{Total ($k=6$)} \\
        \hline
        $128^2$ & 9s & 9s & 6s & 2s & 31s \\
        $256^2$ & 38s & 61s & 38s & 31s & 228s \\
        $512^2$ & 245s & 294s & 240s & 194s & 1303s \\
        $1024^2$ & 1598s & 2095s & 1586s & 1489s & 9077s \\
        \hline
      \end{tabular}
    \end{center}
  \end{minipage}
\end{table*}

\subsection{Algorithm}

The procedure consists in solving a sequence of FF problems that quickly
converges to a solution of the NF one.
Details can be found in Algorithm~\ref{algo:near-field}.
In this algorithm, \texttt{SOLVE\_FF($ T, \rho,
Y, \sigma, \eta$)} denotes an algorithm that solves the FF problem
between a source $ \rho : T \to \Rsp^+ $ and a target $
\sigma=\sum_i\sigma_i\delta_{y_i} $ supported on $ Y \subset \Sph^2 $ for a numerical
error $ \eta $. It can for instance be Step 2 of Algorithm~\ref{algo:main}.
This algorithm is used to produce all the lenses and mirrors of the article,
except for the first image of Figure~\ref{fig:pillows}. 

\subsection{Convergence analysis}

\begin{figure}
    \centering
    \includegraphics[width=.7\linewidth]{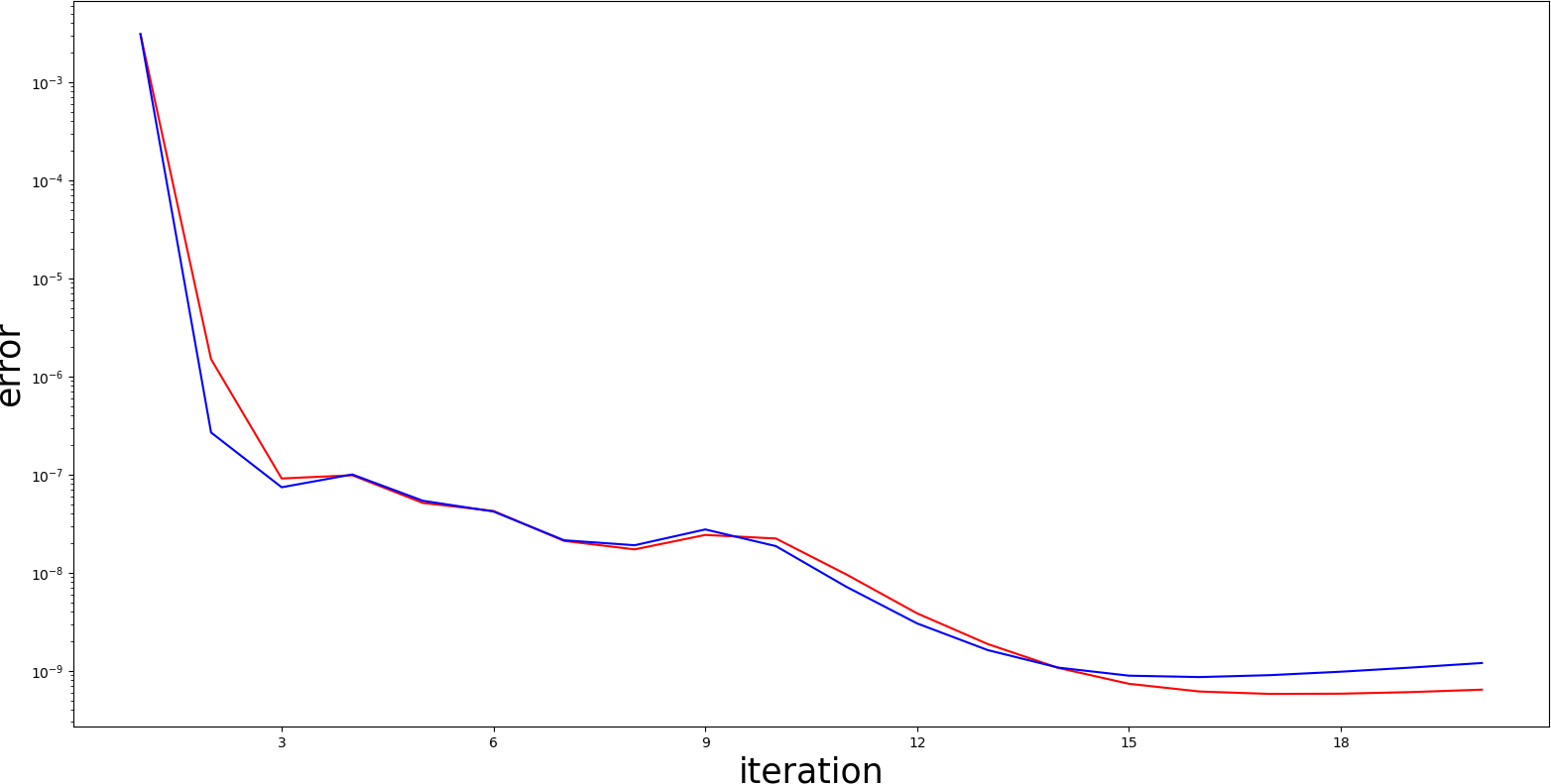}

    \caption{Mean error $ (\nr{c^{k+1} - c^{k}}_1/N) $ in
        Algorithm~\ref{algo:near-field} for a uniform collimated light
        source. Red: \cslens with the
        \textsc{Train} target; Blue: \csmirror with the
        \textsc{Train} target.
        The y-axis is in logarithmic scale.
    }
    \label{fig:near-field-errors}
\end{figure}

It is clear that when a fixed point is reached in Algorithm~\ref{algo:near-field}, the corresponding weight
vector $ \psi $ is a \jocelynbis{good approximation} of the NF problem.
In practice, the algorithm converges very quickly:
in all our examples, after 6 iterations, we get an error $\eta_{NF} $ of $ 10^{-6}$, see Figure~\ref{fig:near-field-errors}. 

\begin{figure}
    \centering


    \includegraphics[width=.23\linewidth]{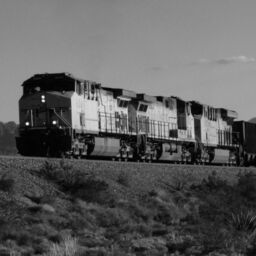}
    \includegraphics[width=.23\linewidth]{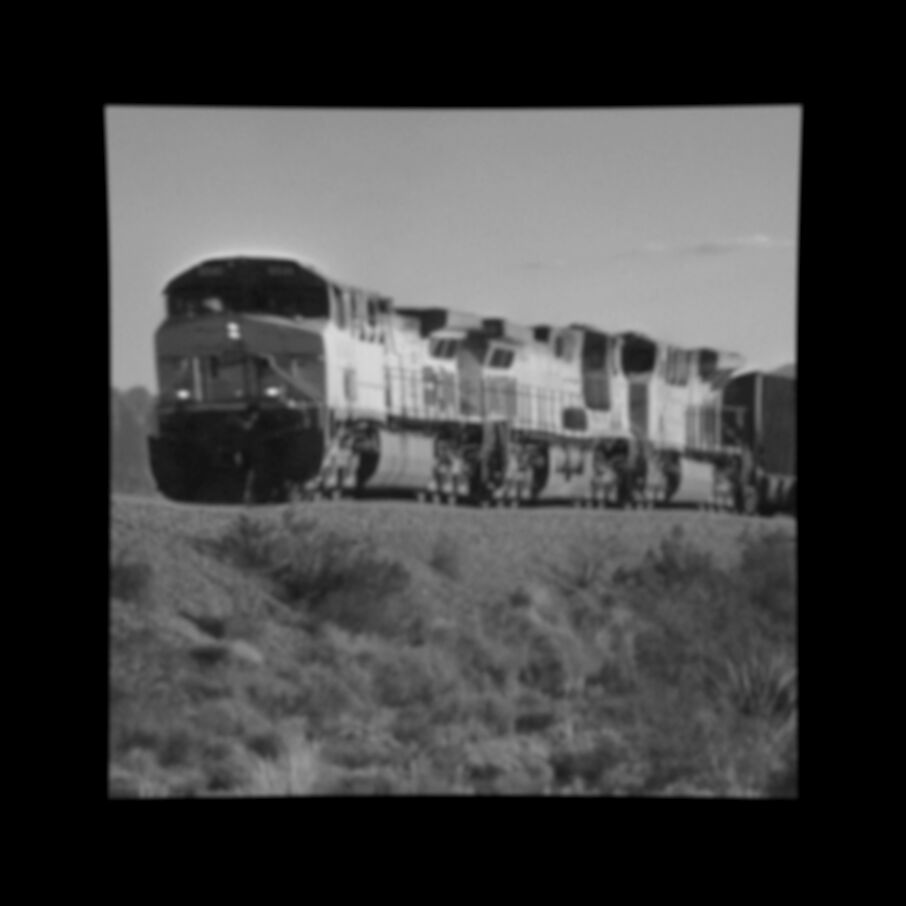}
    \includegraphics[width=.23\linewidth]{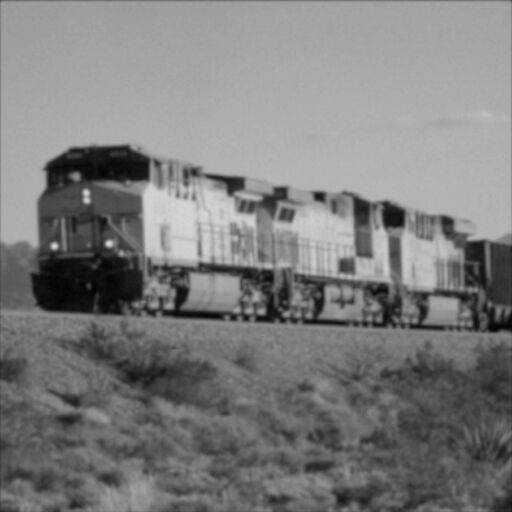}
    \includegraphics[width=.23\linewidth]{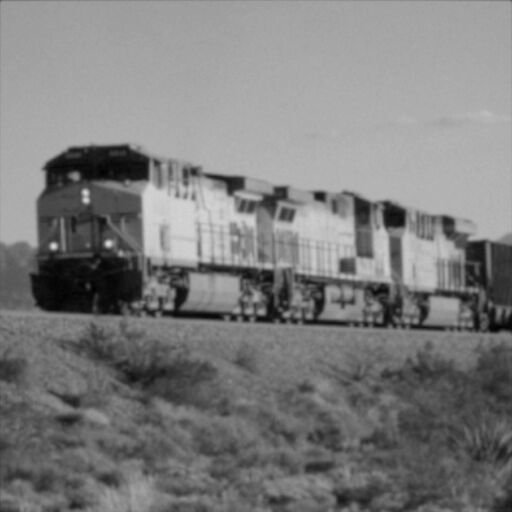}


    \includegraphics[width=.23\linewidth]{new/luxrender/train/train_original.jpg}
    \includegraphics[width=.23\linewidth]{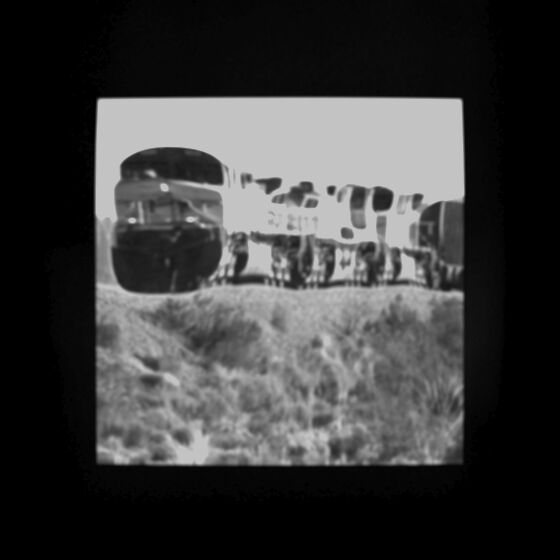}
    \includegraphics[width=.23\linewidth]{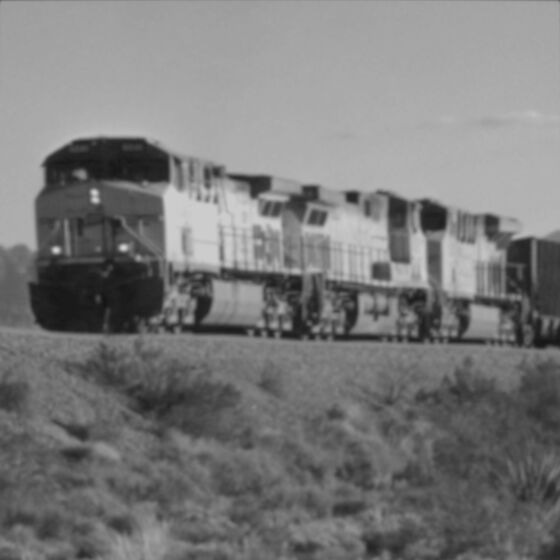}
    \includegraphics[width=.23\linewidth]{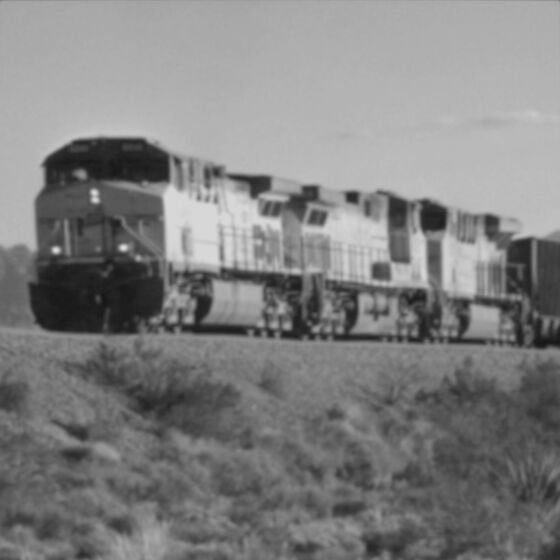}

    \caption{Forward simulation  of different iterations of
        Algorithm~\ref{algo:near-field}.
        From left to right: target image, $1^{st}$ iteration; $ 2^{nd} $
        iteration; after 6 iterations. First row: Concave \cslens
        \textsc{Train}. Second row: Convex \csmirror \textsc{Train}.}
    \label{fig:near-field-iterative-renderings}
\end{figure}

\jocelynbis{
We have no guarantee on the convergence of the discretization. However, note that the set of reflected (or refracted) rays emanating from a visibility cell has a diameter proportional to the diameter of the visibility cell. This generates a blur whose diameter is also proportional to the diameter of the cells. In practice, we observe that these cells have a small diameter (see Figures~\ref{fig:diagrams-problems} and \ref{fig:res-non-uniform}) and indeed the blur is unnoticeable.  
}
The convergence is illustrated in Figure~\ref{fig:near-field-iterative-renderings}
where we show LuxRender renderings at iteration $ 1 $ (which corresponds to the
FF setting), at iteration $ 2 $ and after $6$ steps (when the error
\jocelynbis{$\|{c^{k+1} - c^{k}}\|_1/N$ becomes smaller than $ \eta_{NF} = 10^{-6} $}).
In all the images, we look at the projection of the refracted illumination onto the target
screen. One can see that the first one is distorted since it corresponds to the
projection of an image supported on the sphere onto a planar screen. Starting from
the second iteration of the algorithm, the image is not distorted anymore.

The convergence of this algorithm is also illustrated in
Figure~\ref{fig:pillows}. On the first image of
Figure~\ref{fig:pillows}, the optical component is composed of three
lenses that solve the FF problem. We observe that the LuxRender
rendering creates three shifted copies of the same image. To be more
precise, the translation between the 3 images is \jocelynbis{the same
  as the one between the lenses i.e. if each lens has a width of $1$
  then each image is also shifted by $1$}.
In the second image, the nine lenses solve the NF problem for an image at a finite distance. The fact that the images do superimpose show that the NF problem is accurately solved. The quality of the first image of Figure~\ref{fig:teaser} also assesses that the NF problem is solved accurately.


\subsection{Performance}

We report performance of algorithms~\ref{algo:main} and \ref{algo:near-field} to solve the \csmirror case for
the \textsc{Train} target on a laptop with an i7 CPU.
Table~\ref{tbl:ff-running-times} shows the running times in the FF setting,
We underline that the number of iterations in the Newton step remains
low: in all our examples it varies from $ 10 $ to $20$. This means
that the computational cost of the method is concentrated in the
computation of the functional $G$, its Jacobian matrix $\D G$ and the
resolution of the linear system.  We believe that there is much room
for improvement in the first two steps, by optimizing the computation
of visibility cells, and by using an explicit computation of $\D G$
instead of relying on automatic differentiation.


In the NF setting, with the same configuration, the results are summarized in
Table~\ref{tbl:nf-running-times}. The total running time is the one after $k=6$ iterations.
Note that the running time of the second iteration is greater than the first
one. Indeed, in Algorithm~\ref{algo:near-field}, we saw that we use the
weights found at one step to initialize the next one. Since the target directions
change greatly between the first and second steps, these weights
do not provide a good initialization. To fix that,
we automatically apply a simple perturbation to obtain good initial weights.
Starting from the third step, the running times decrease as the target
directions do not move a lot at each iteration.

\section{Results and Discussion}
\label{sec:results}
In this section, we present several numerical examples for the
different problems previously described as well as some other
applications.  In the experiments, we take $\kappa=1.5$. Unless stated
otherwise, the light source is chosen uniform and the discretization of
the target (number of Diracs $ N $) is equal to the size of the
image. The stopping criterion to Newton's algorithm (Algorithm~\ref{algo:newton}) is set to $\eta = 10^{-8}$.
\jocelyn{Since the convergence of Algorithm~\ref{algo:near-field} is always
    fast, we do not use here the stopping criteria $\eta_{NF}$ and stop the algorithm after $k=6$ iterations in all
    the examples presented hereafter.}

\subsection{Evaluation strategy}



The output of our algorithm is a triangulation equipped with a normal at each
vertex.
\jocelyn{In all the simulations, we use the \textbf{LuxRender} rendering
engine, with Bidirectional Path Tracing
 combined with a Sobol sampler and the Fresnel
coefficient is not taken into account.


}

\subsection{Results}

\begin{figure}
\begin{center}
    \includegraphics[width=.34\linewidth]{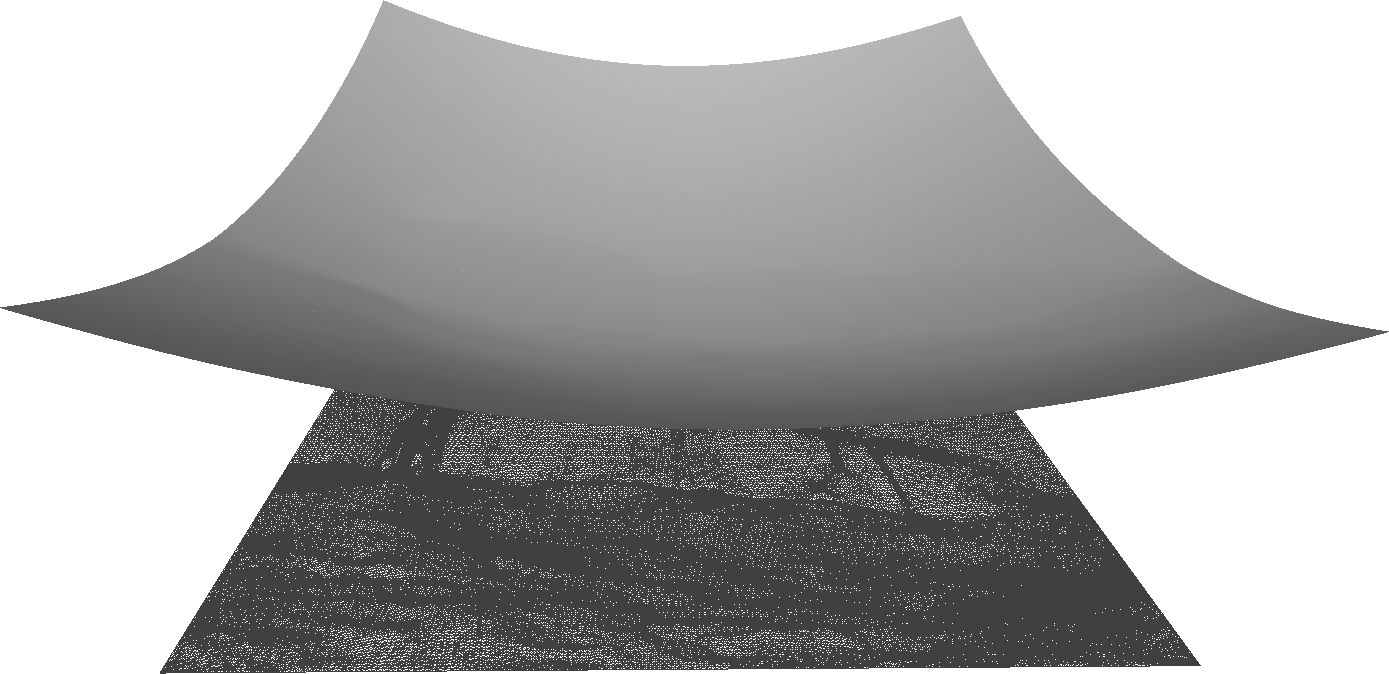}
    \includegraphics[width=.34\linewidth]{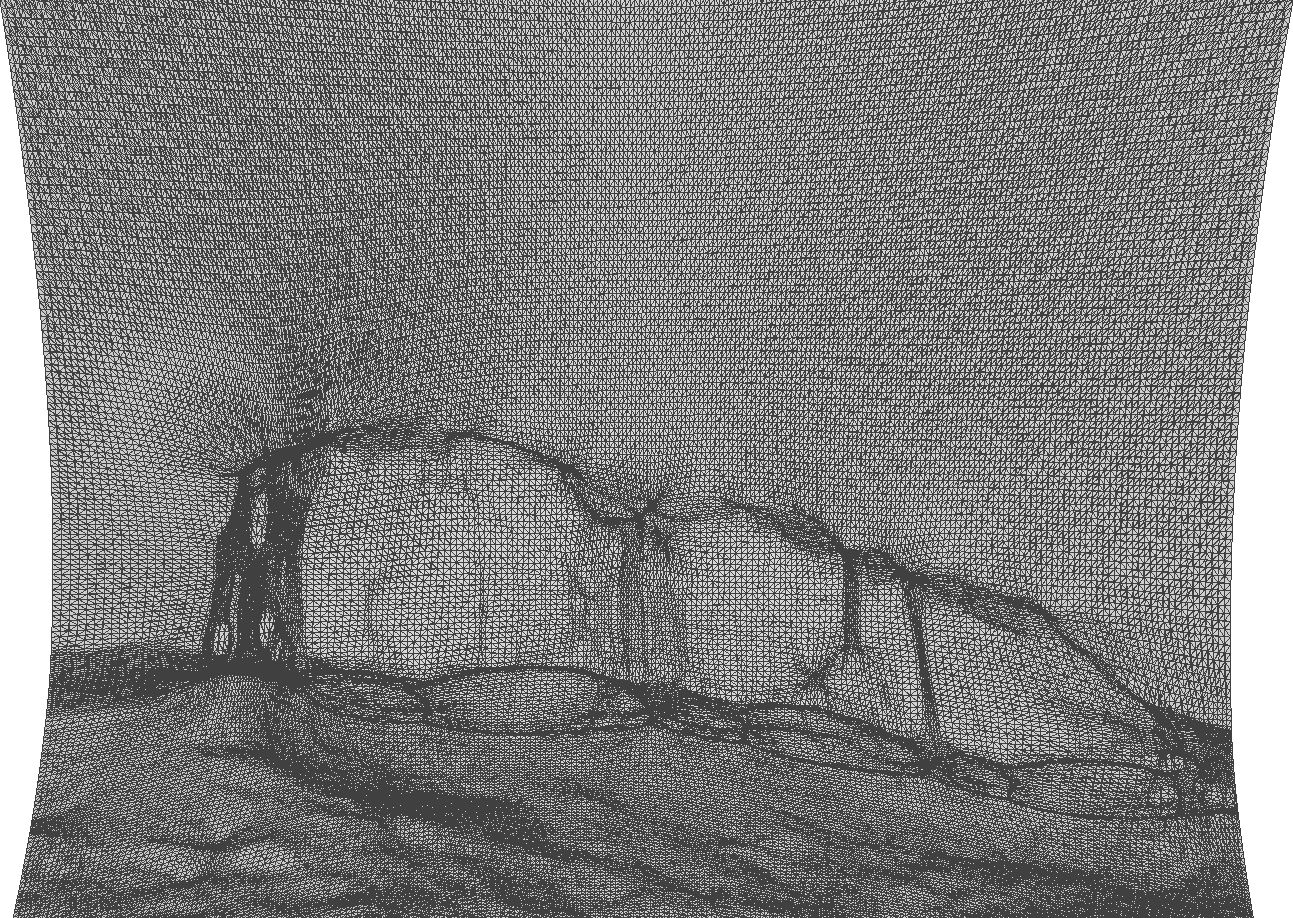}
    \includegraphics[width=.28\linewidth]{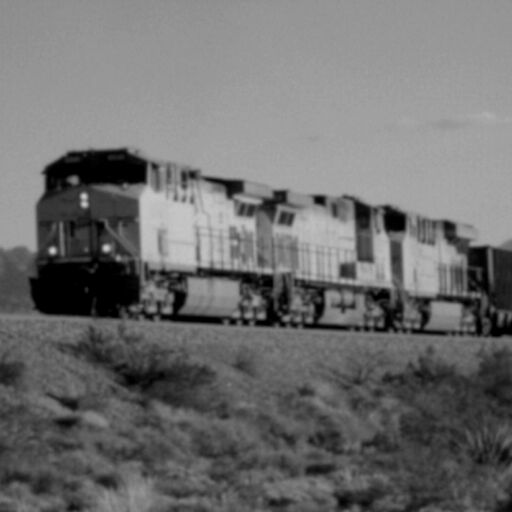}
\end{center}
\begin{center}
    \includegraphics[width=.34\linewidth]{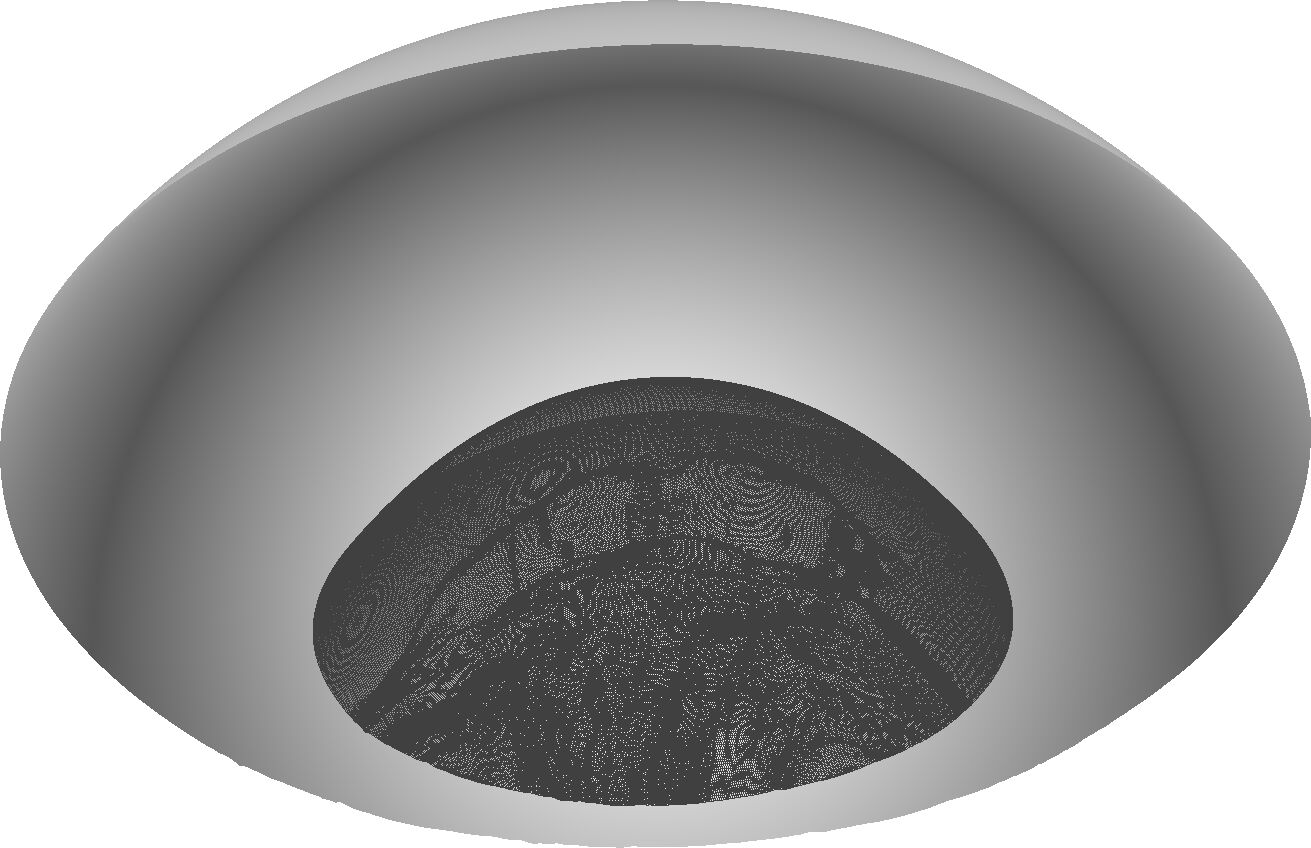}
    \includegraphics[width=.34\linewidth]{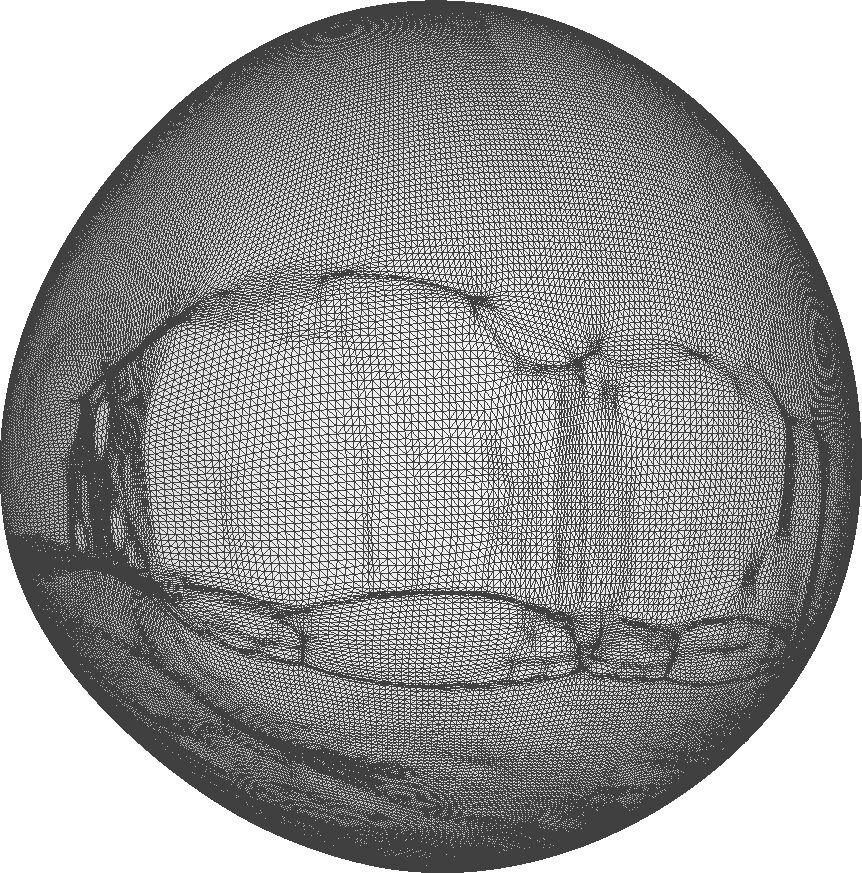}
    \includegraphics[width=.28\linewidth]{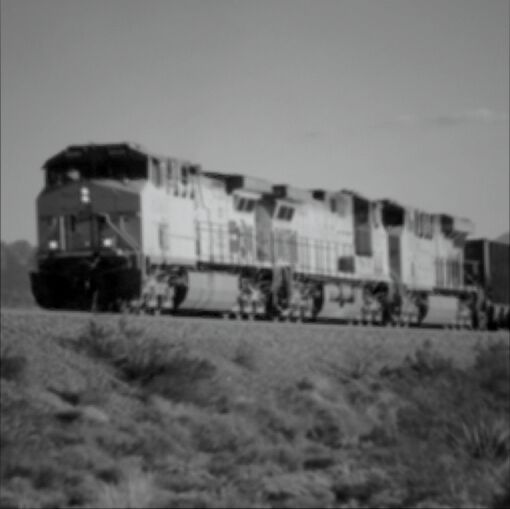}
\end{center}
\begin{center}
    \includegraphics[width=.34\linewidth]{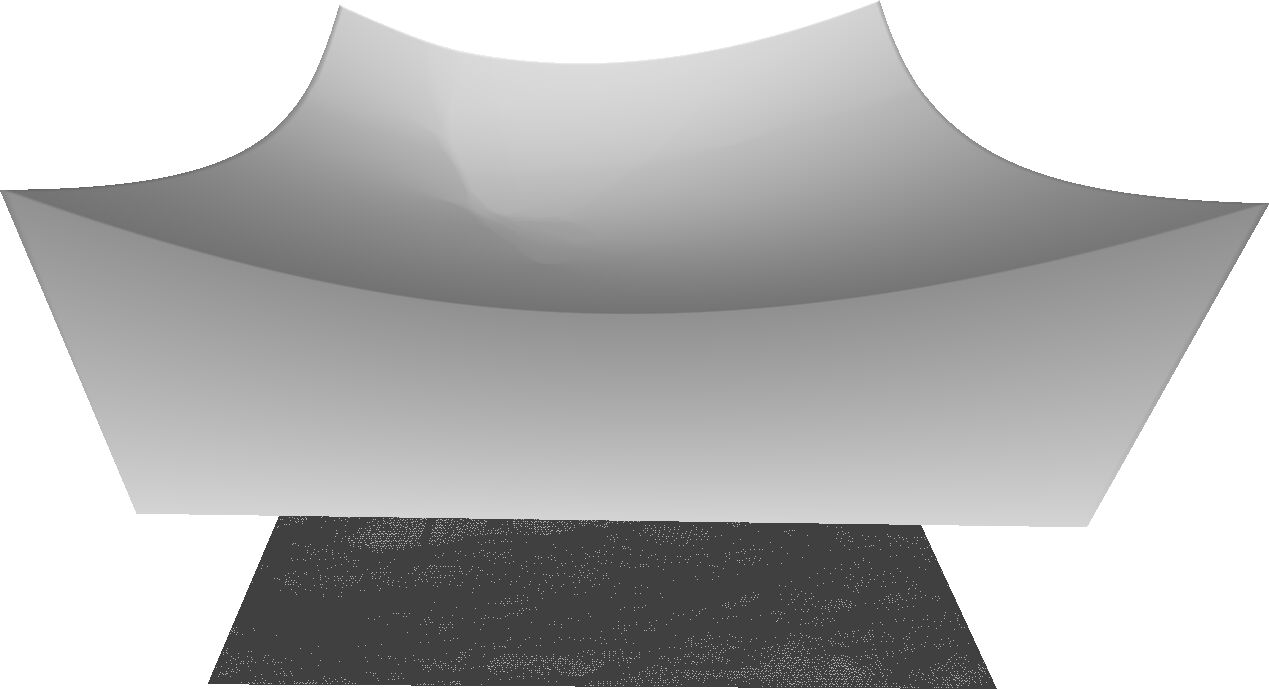}
    \includegraphics[width=.34\linewidth]{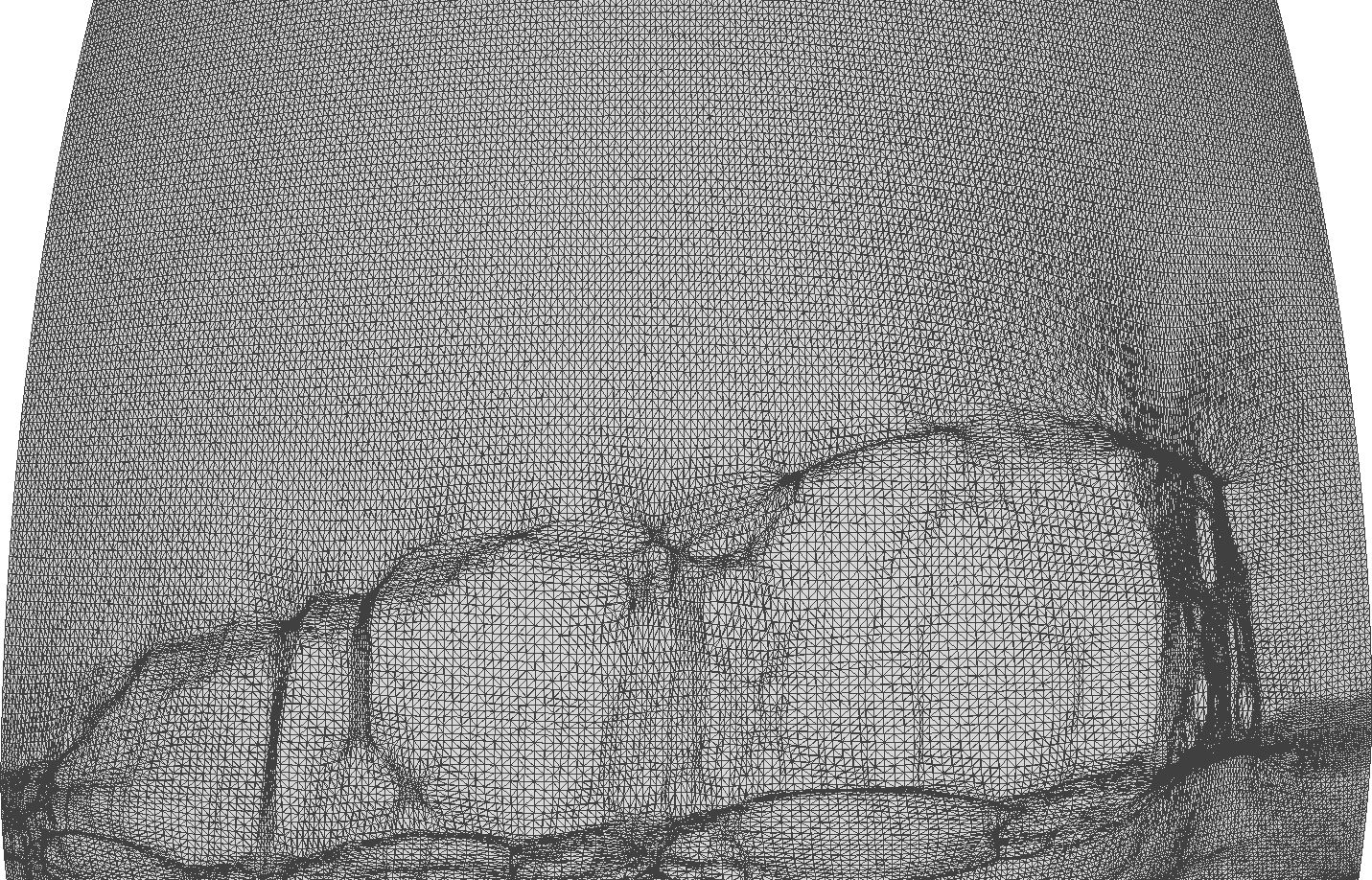}
    \includegraphics[width=.28\linewidth]{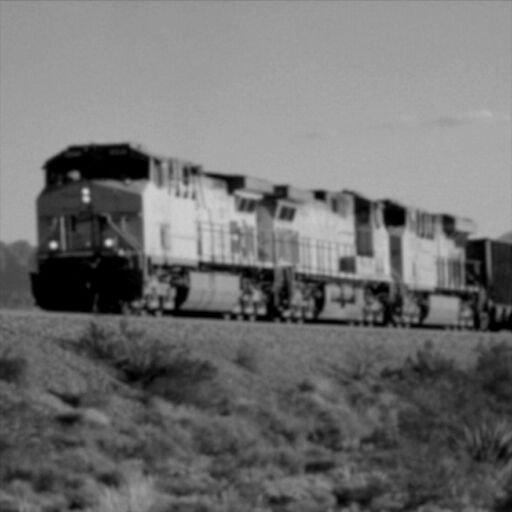}
\end{center}
\begin{center}
        \includegraphics[width=.32\linewidth]{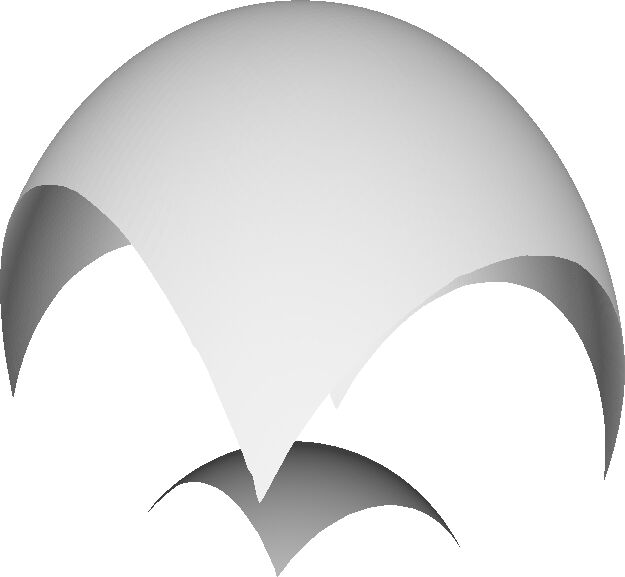}
        \includegraphics[width=.34\linewidth]{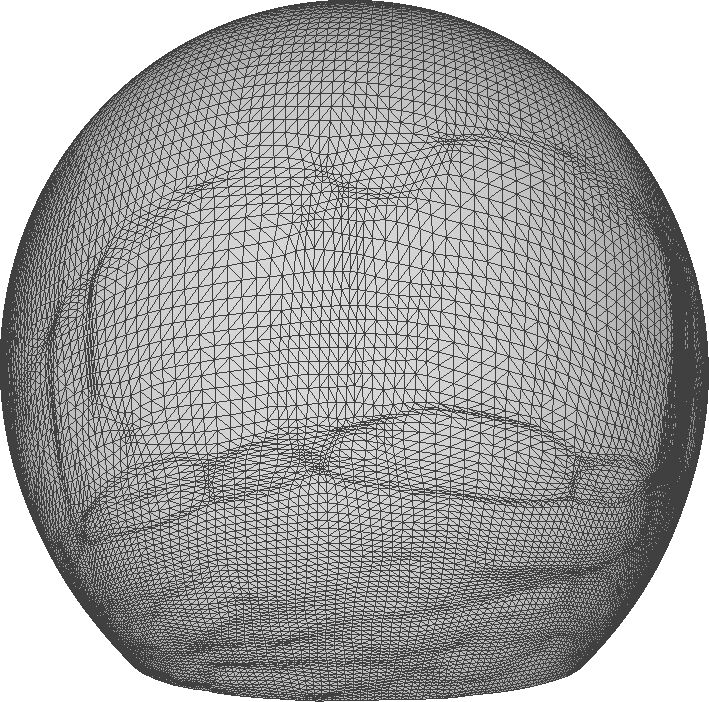}
        \includegraphics[width=.28\linewidth]{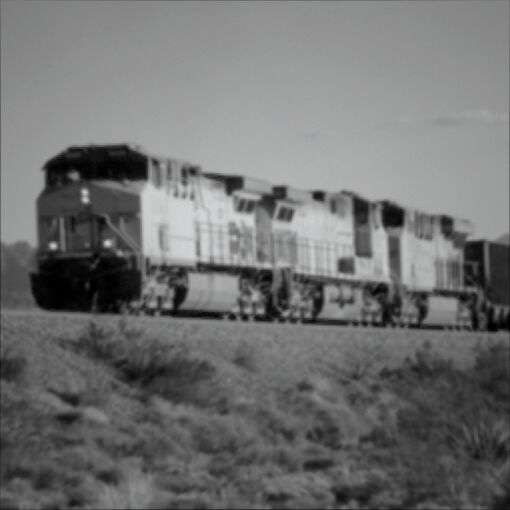}
\end{center}

\caption{\textbf{Four non-imaging problems solved with Algorithm~\ref{algo:main}}.
      \label{fig:diagrams-problems}\textit{From left to right:} visibility diagram on $ X_\rho $
        (wireframe) with the optical component $\Ref$, Triangulation $ \Ref_T $  of $\Ref$;
        \jocelyn{forward simulation using LuxRender.}
        \textit{From top to bottom:} Convex Collimated Source Mirror; Concave Point Source Mirror;
        Concave Collimated Source  Lens; Point Source Lens (union of ellipsoids).
}
\end{figure}

\begin{figure}

    \begin{center}
        \includegraphics[width=.26\linewidth]{new/luxrender/train/train_original.jpg}
        \includegraphics[width=.26\linewidth]{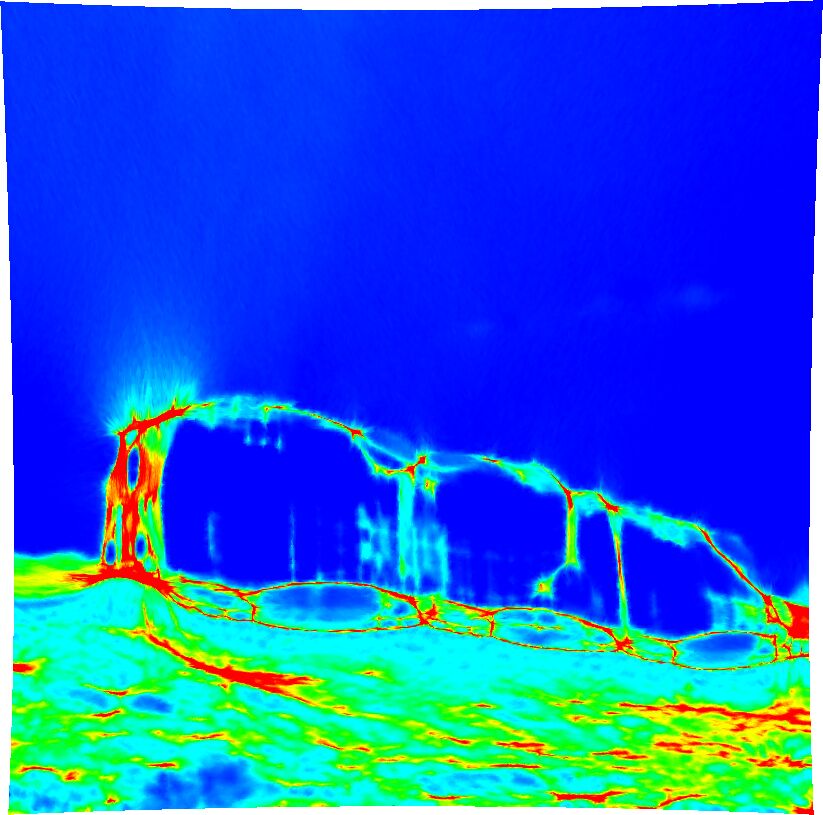}
        \includegraphics[width=.26\linewidth]{new/luxrender/train/cs_mirror_train_256_render_fixed.jpg}
    \end{center}%


    \begin{center}
        \includegraphics[width=.26\linewidth]{images/results/hikari_300x300_original_white}
        \includegraphics[width=.26\linewidth]{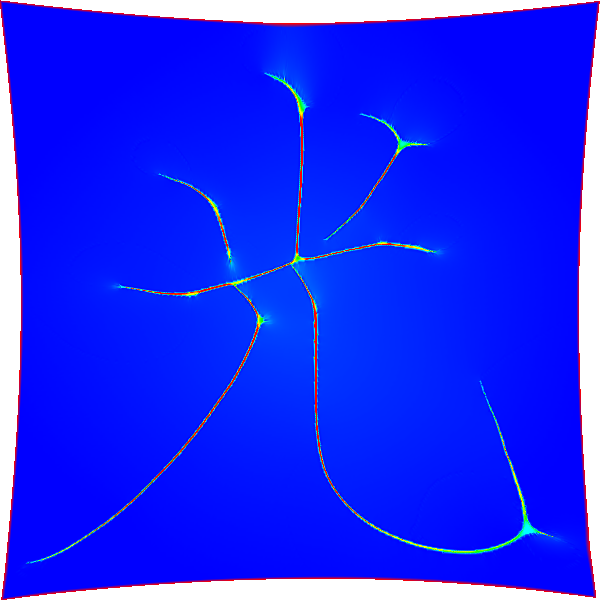}
        \includegraphics[width=.26\linewidth]{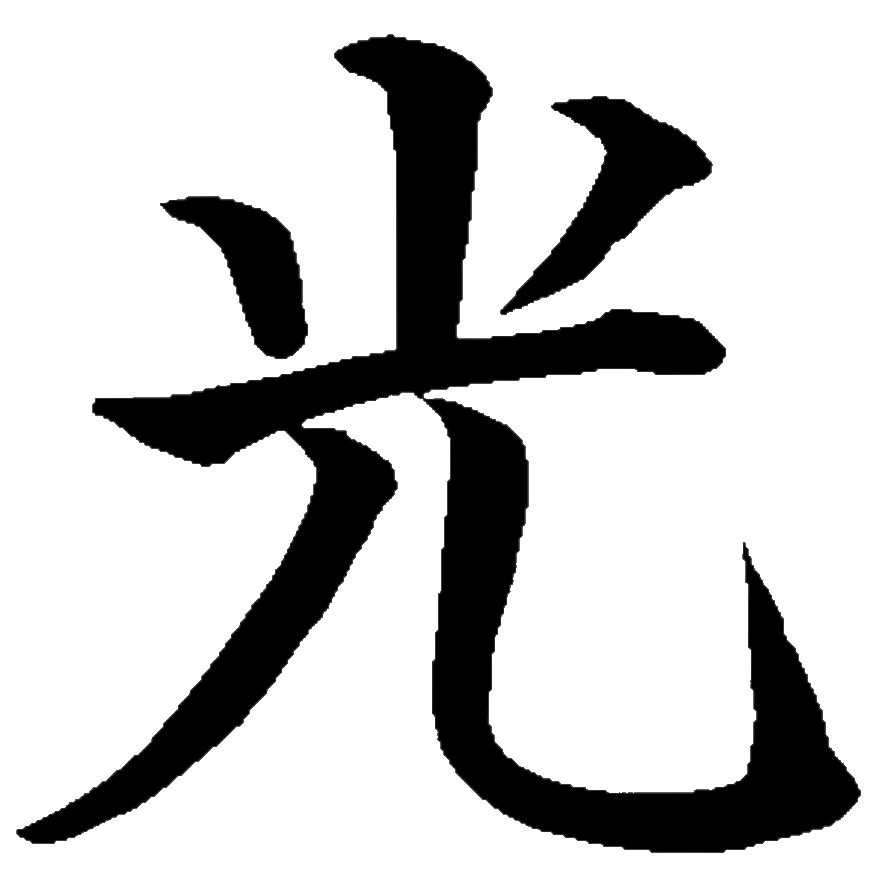}
    \end{center}%

    \begin{center}
        \includegraphics[width=.26\linewidth]{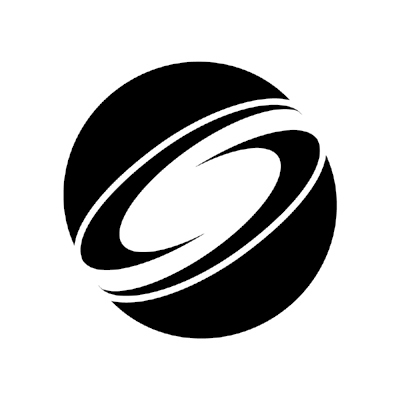}
        \includegraphics[width=.26\linewidth]{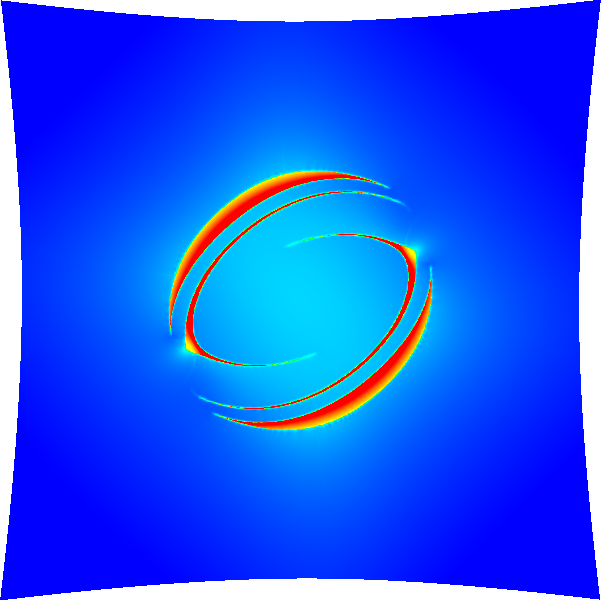}
        \includegraphics[width=.26\linewidth]{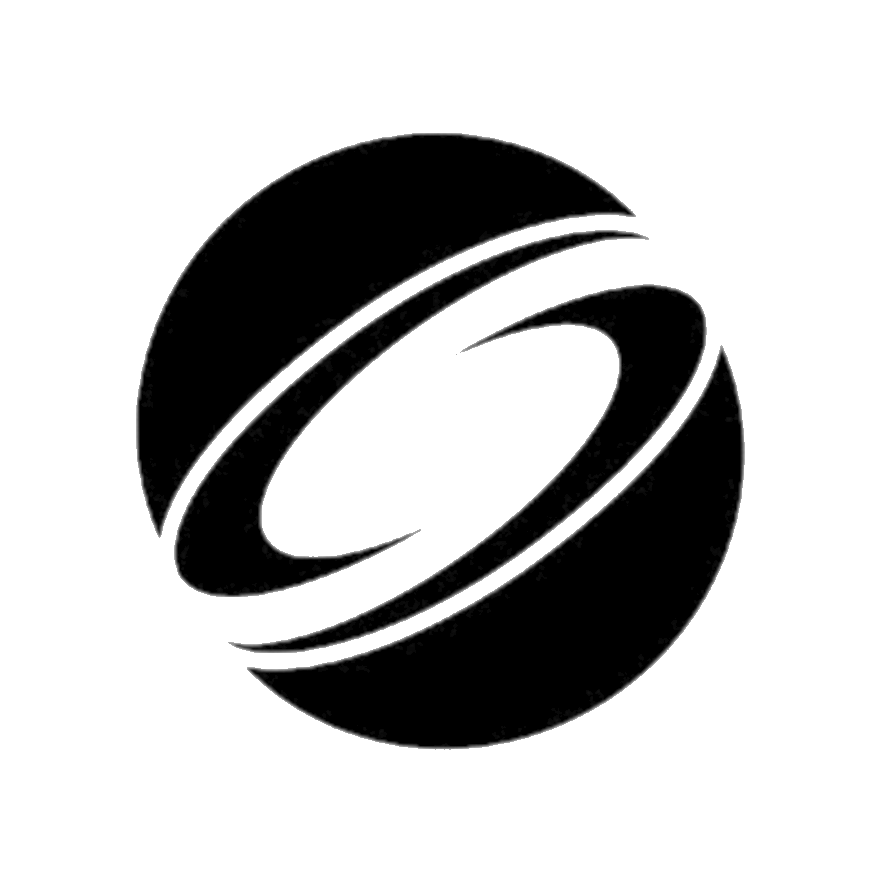}
    \end{center}%

    \caption{\textbf{Convex Collimated Source Mirror} problem with a uniform light source for different
        target distributions.
        \textit{From left to right: } target distribution,
        mean curvature of the mirror, forward simulation \jocelyn{using
            LuxRender}.
        Dimensions of images (top to bottom): 256$^2$, 300$^2$,
        400$^2$.
    }
    \label{fig:res-csff}
\end{figure}

\begin{figure}

    \begin{center}
        \includegraphics[width=.26\linewidth]{new/luxrender/train/train_original.jpg}
        \includegraphics[width=.26\linewidth]{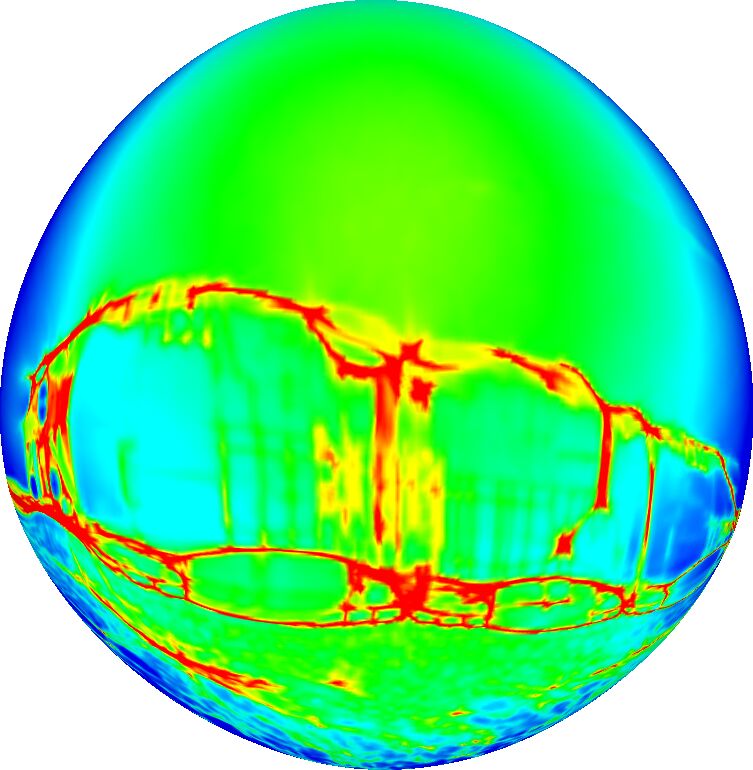}
        \includegraphics[width=.26\linewidth]{new/luxrender/train/ps_mirror_render_fixed.jpg}
    \end{center}%


    \begin{center}
        \includegraphics[width=.26\linewidth]{images/results/hikari_300x300_original_white}
        \includegraphics[width=.26\linewidth]{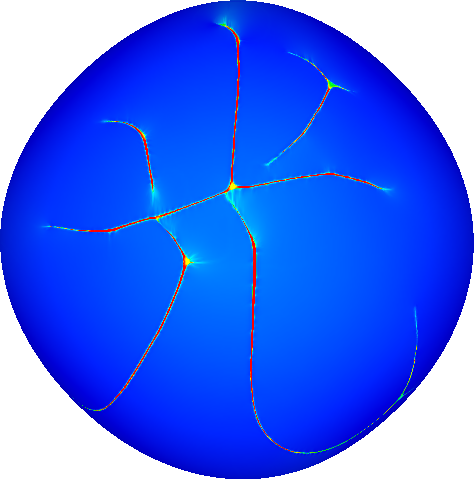}
        \includegraphics[width=.26\linewidth]{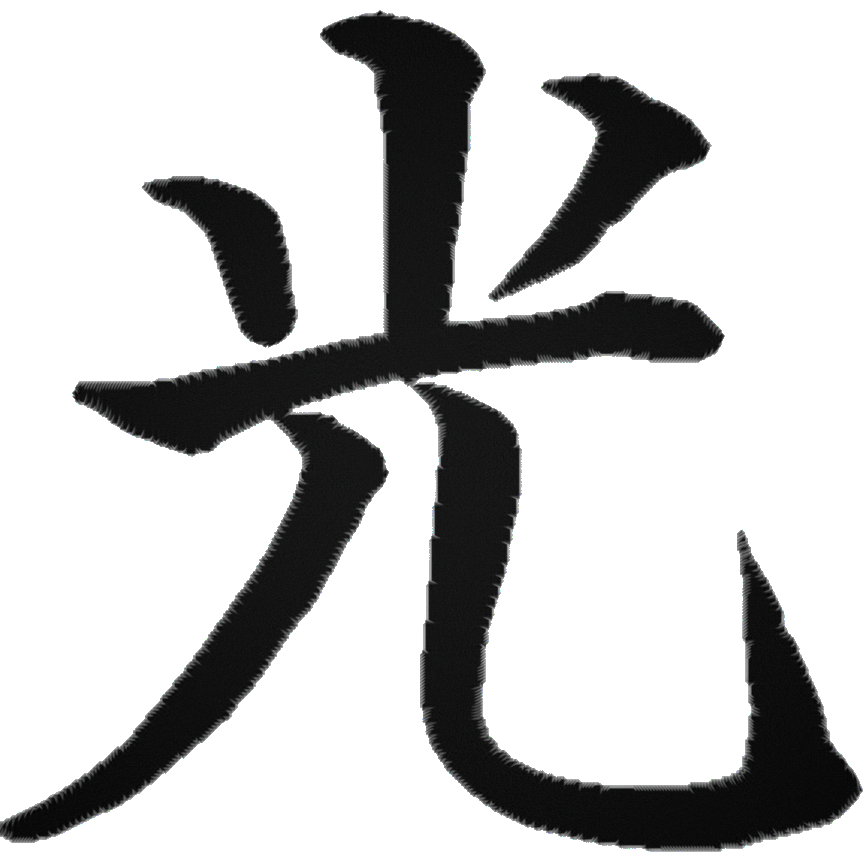}
    \end{center}%
    \begin{center}
        \includegraphics[width=.26\linewidth]{images/results/logo_siggraph_400x400_original_black_on_white}
        \includegraphics[width=.26\linewidth]{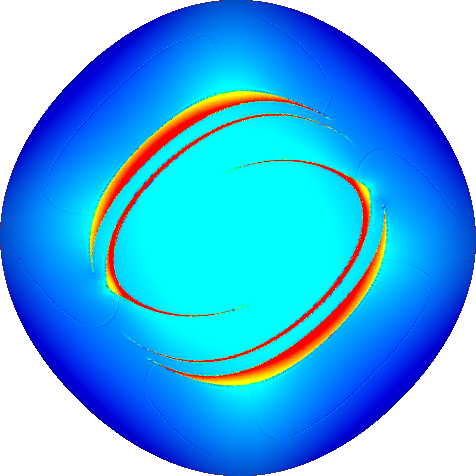}
        \includegraphics[width=.26\linewidth]{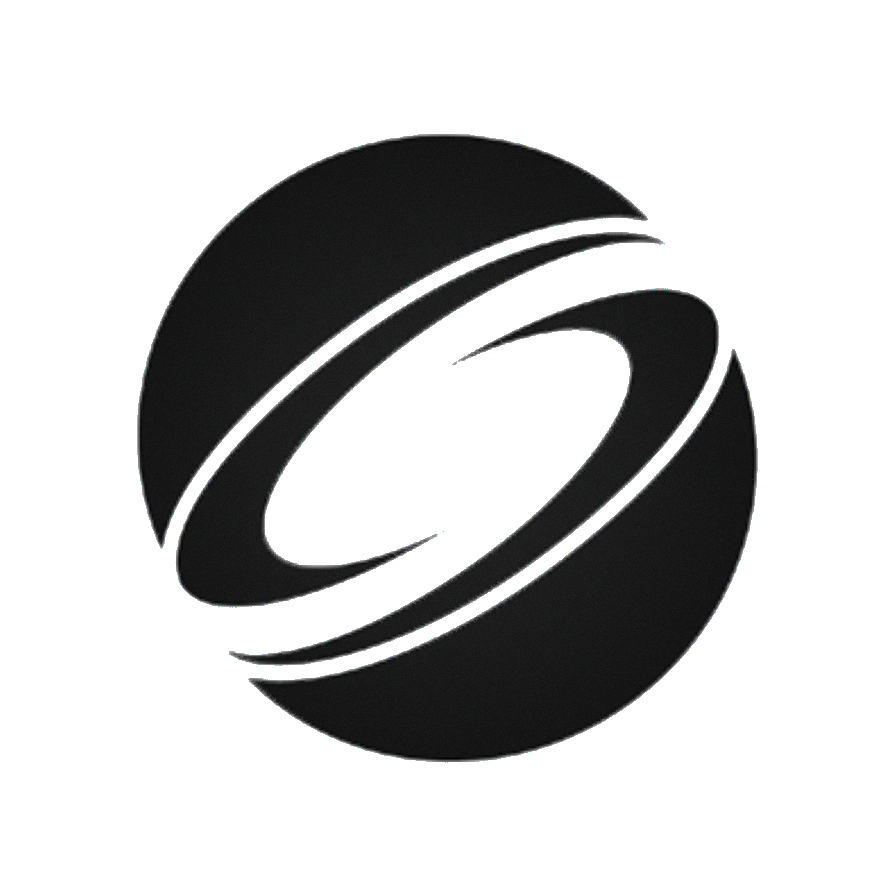}
    \end{center}%
    \caption{\textbf{Concave Point Source Mirror} problem for a uniform point light source with
        different target distributions.
        \textit{From left to right:} target distribution, mean curvature of the
        mirror (top view), forward simulation
        \jocelyn{using LuxRender.}
        Dimensions of images (top to bottom): $256^2$, 300$^2$, 400$^2$. 
        \label{fig:res-psff}}
\end{figure}

\paragraph{Genericity.}
Our algorithm is able to solve eight different optical component design
problems. We present for instance in Figure~\ref{fig:diagrams-problems} four
examples for which we display the visibility diagram of $X_\rho$ as well as the
optical component (lens or mirror) above it, a mesh of the optical component and
a forward simulation using \jocelyn{LuxRender}.

\begin{figure}

    \begin{center}
        \includegraphics[width=.26\linewidth]{new/luxrender/train/train_original.jpg}
        \includegraphics[width=.26\linewidth]{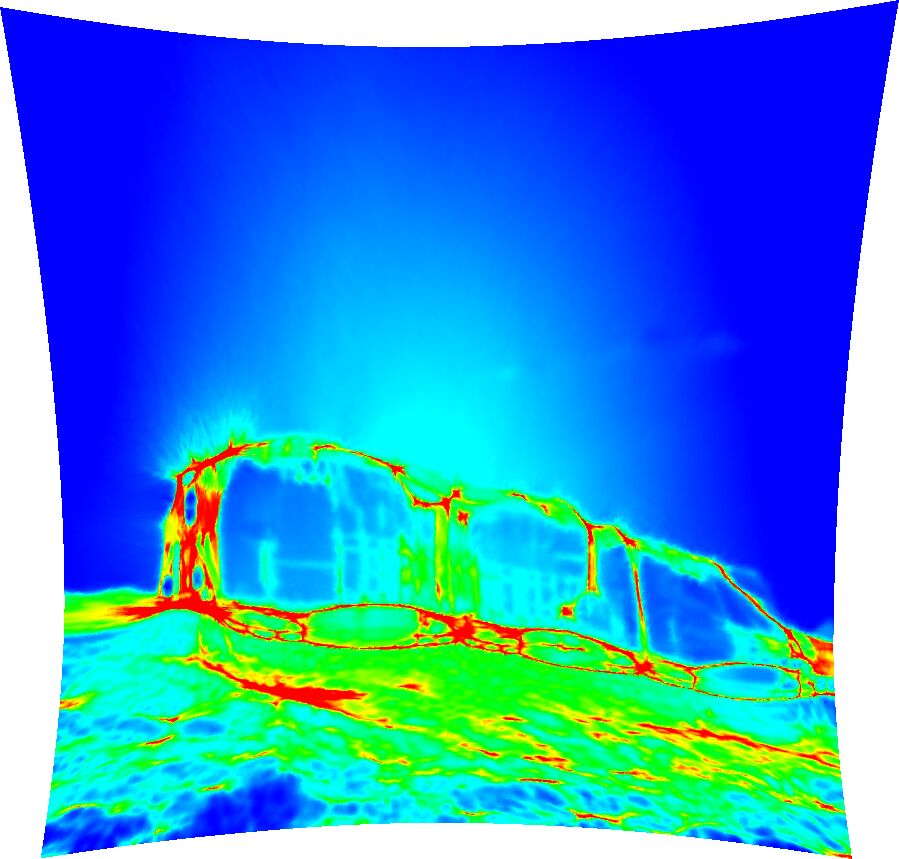}
        \includegraphics[width=.26\linewidth]{new/luxrender/train/cs_lens_train_256_render_fixed.jpg}
    \end{center}%


    \begin{center}
        \includegraphics[width=.26\linewidth]{images/results/hikari_300x300_original_white}
        \includegraphics[width=.26\linewidth]{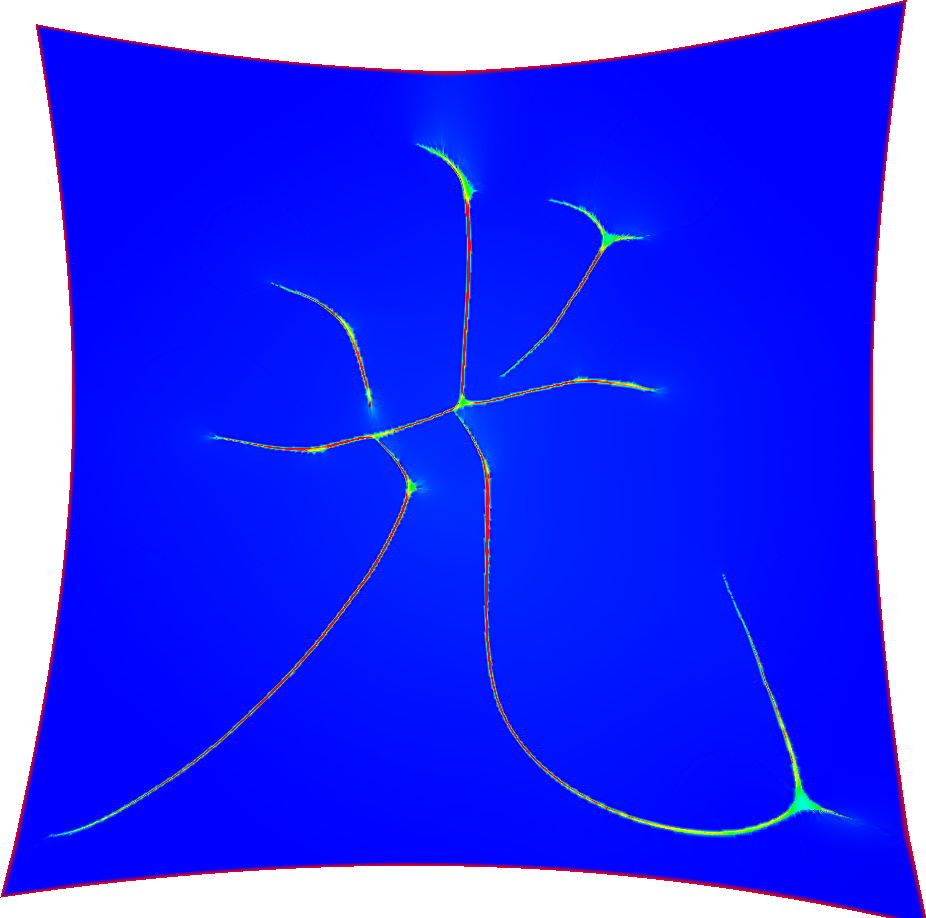}
        \includegraphics[width=.26\linewidth]{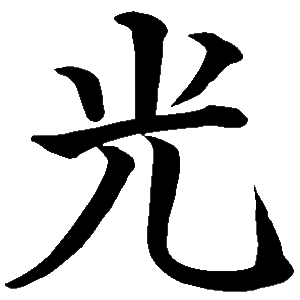}
    \end{center}%


    \begin{center}
        \includegraphics[width=.26\linewidth]{images/results/logo_siggraph_400x400_original_black_on_white}
        \includegraphics[width=.26\linewidth]{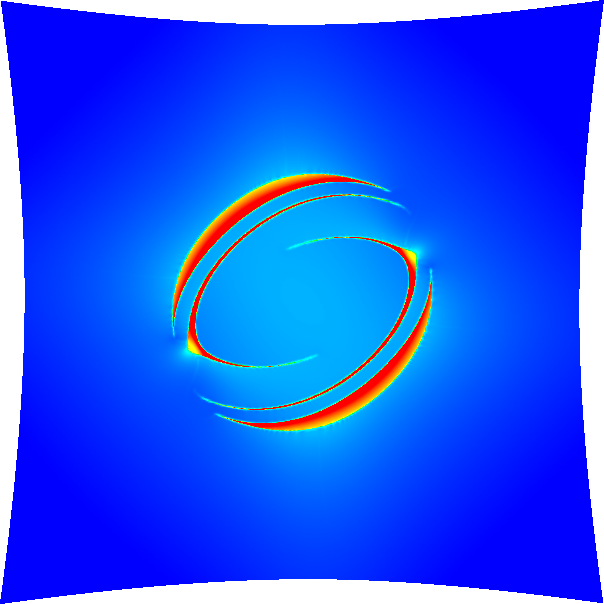}
        \includegraphics[width=.26\linewidth]{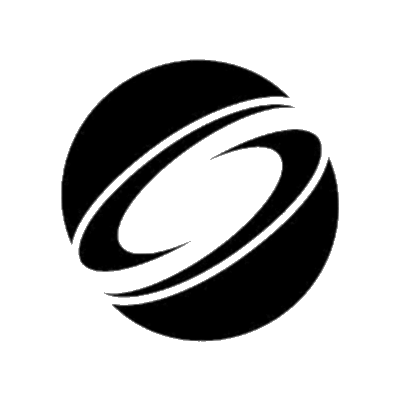}
    \end{center}%

    \caption{\textbf{Concave Collimated Source Lens} with a uniform light source for different
        target distributions.
        \textit{From left to right:} target distribution, mean curvature
         of the lens (top view), forward simulation \jocelyn{using
            LuxRender}.
        Dimensions of images (top to bottom): 256$^2$, 300$^2$,
        400$^2$. \vspace{-.5cm}}
    \label{fig:res-cs_lens}
\end{figure}
\begin{figure}

    \begin{center}
        \includegraphics[width=.26\linewidth]{new/luxrender/train/train_original.jpg}
        \includegraphics[width=.26\linewidth]{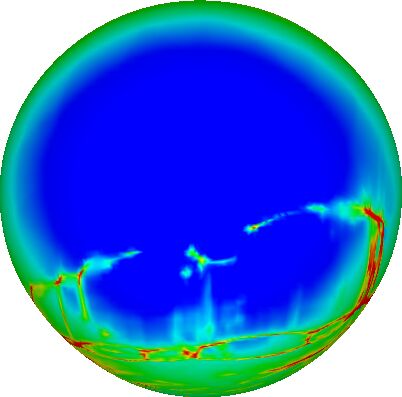}
        \includegraphics[width=.26\linewidth]{new/luxrender/train/ps_lens_render_fixed.jpg}
    \end{center}%


    \begin{center}
        \includegraphics[width=.26\linewidth]{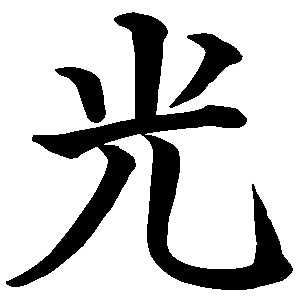}
        \includegraphics[width=.26\linewidth]{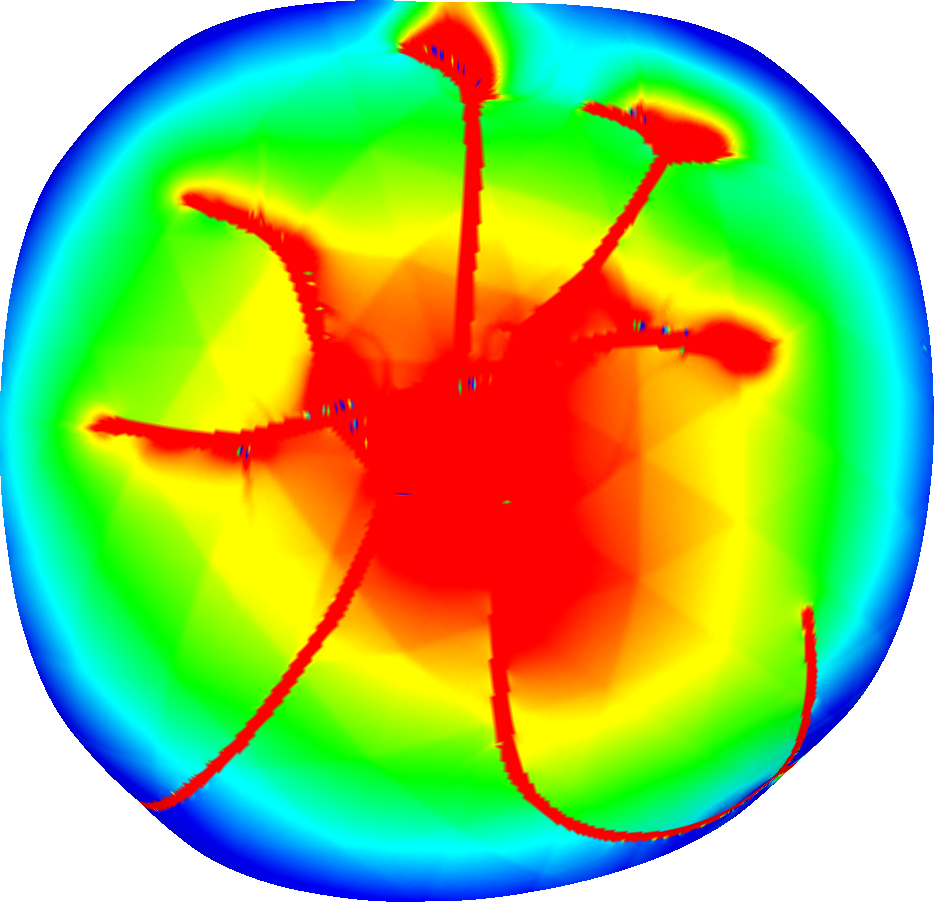}
        \includegraphics[width=.26\linewidth]{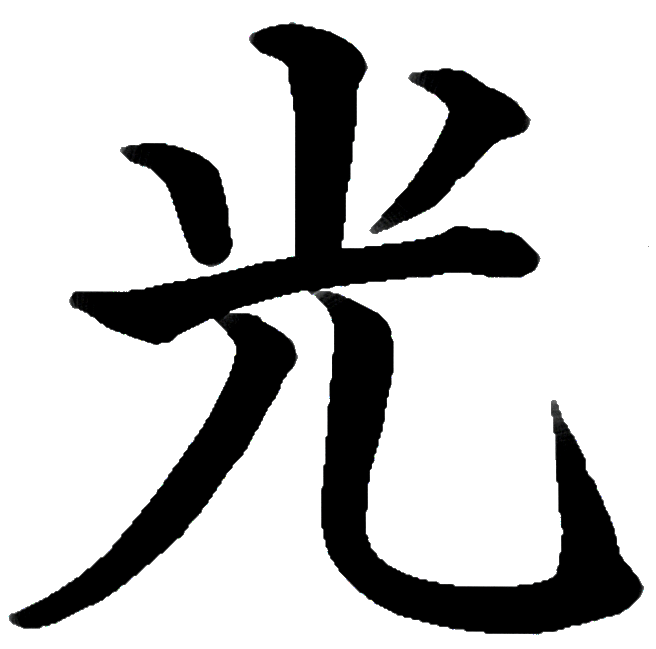}
    \end{center}%

    \begin{center}
        \includegraphics[width=.26\linewidth]{images/results/logo_siggraph_400x400_original_black_on_white}
        \includegraphics[width=.26\linewidth]{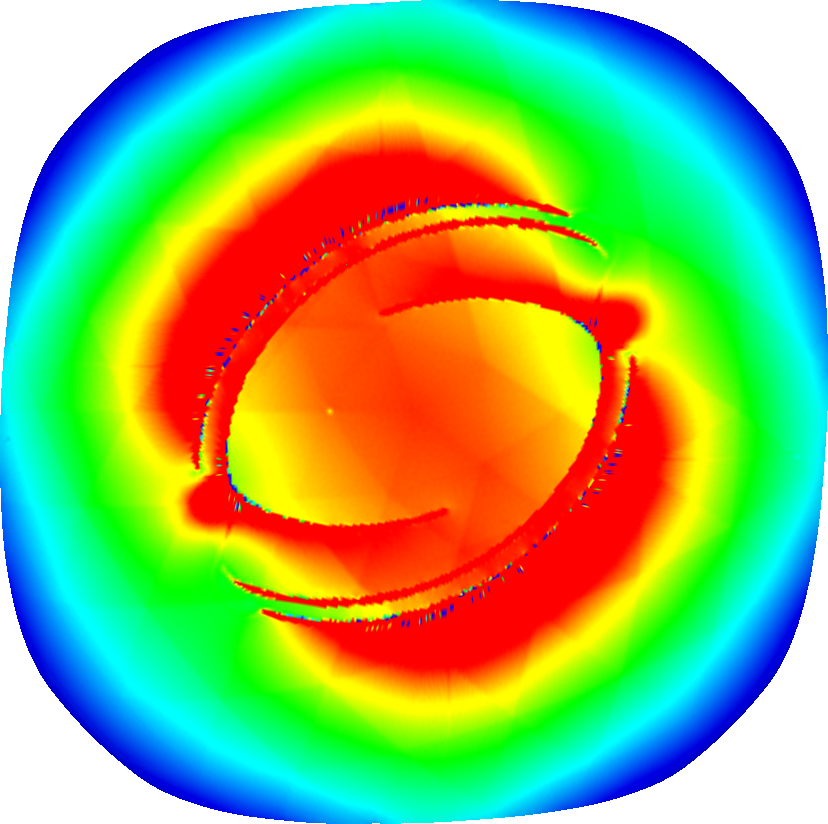}
        \includegraphics[width=.26\linewidth]{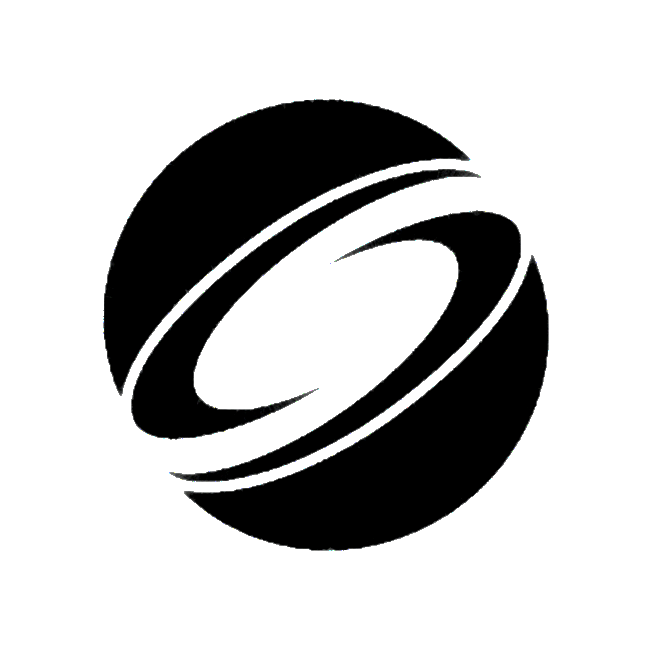}
    \end{center}%


    \caption{\textbf{Point Source Lens} with a uniform light source for different target
        distributions. The lens surface is the boundary of the union of filled ellipsoids, hence is not convex, nor concave.
        \textit{From left to right: } target distribution, mean
        curvature of the lens (top view), forward simulation
        \jocelyn{using LuxRender.}
        Dimensions of images (top to bottom): 256$^2$, 300$^2$, \jocelyn{400$^2$}.
     }
    \label{fig:res-ps_lens}
\end{figure}

Then, for the examples of Figures~\ref{fig:res-csff}, \ref{fig:res-psff},
\ref{fig:res-cs_lens}, and \ref{fig:res-ps_lens}, we display the target
distribution as an image; a mean curvature plot (blue represents low mean curvature and red high mean curvature) of the constructed mesh $\Ref_T$ and a forward simulation using
\jocelyn{LuxRender}.

\paragraph{High-contrast and complex target lights.} We can handle any kind of
target distribution.  Figures \ref{fig:res-csff} and \ref{fig:res-psff} shows
several examples of mirror design for respectively a collimated and a
point light source.  Note that we are able to construct mirrors for
smooth images such as the \textsc{Train} image as well as images with totally black
areas (third and fourth rows). We are also able to handle target
supported on \emph{non-convex} sets such as the \textsc{Hikari} and
\textsc{Siggraph} images.  One can notice that since the area of the
visibility cells are equal to the greyscale values of the image then
the triangles have roughly the same size, implying that one can
recognize the target image in the mesh of the surface, see Figure
\ref{fig:diagrams-problems} for zooms on different meshes. The
mean curvature plot shows the discontinuities in the surface which
come from the black areas in the image.  Figures \ref{fig:res-cs_lens}
and \ref{fig:res-ps_lens} show the same kind of results for the lens
design problems \cslens and \pslens.

\begin{figure}
{
    \centering


    \includegraphics[width=.32\linewidth]{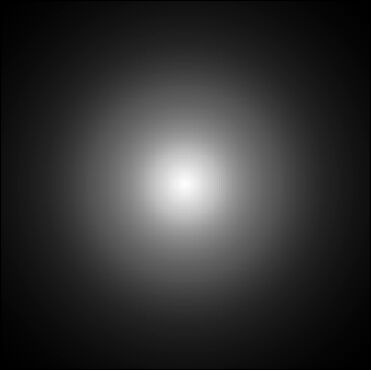}
    \includegraphics[width=.32\linewidth]{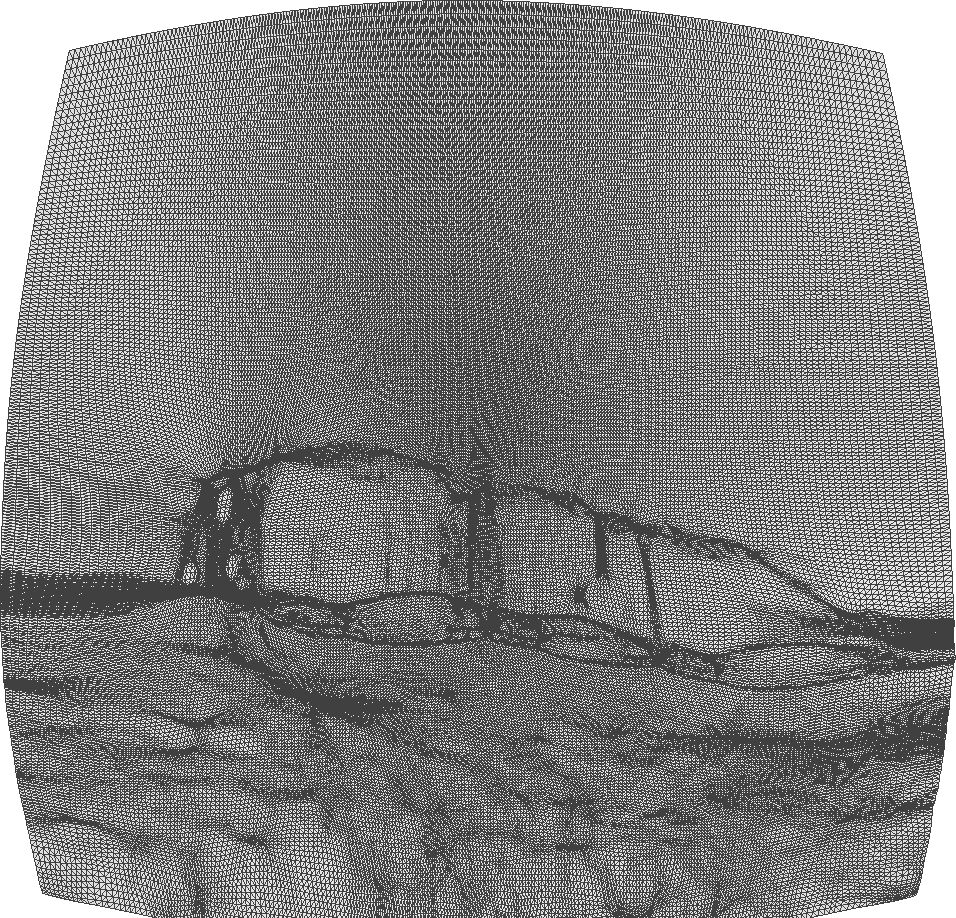}
    \includegraphics[width=.32\linewidth]{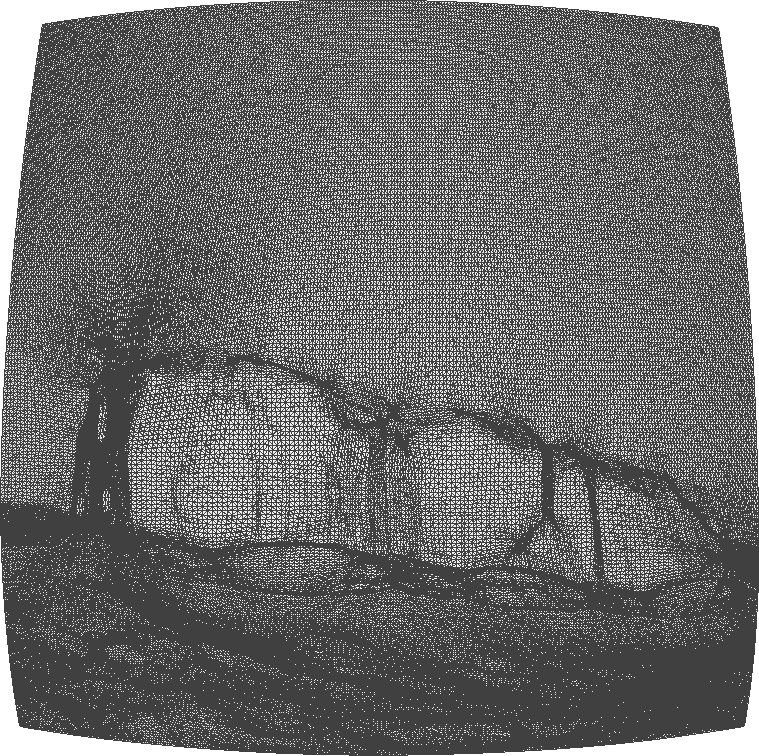}

    \caption{\textbf{Triangulation $\Ref_T$ for a non-uniform light
        source.}  \textit{From left to right}: non uniform collimated
      light source; mesh of the lens for this non-uniform light; mesh
      of the lens for a uniform light
      source.  \label{fig:res-non-uniform}}
    \vspace{-.5cm}
}
\end{figure}

\paragraph{Non-uniform light sources.}
Our algorithms can be used with non-uniform light sources.  Below, we
compare the meshes that are generated in the \cslens case when the
source is either uniform or a Gaussian (Fig. \ref{fig:res-non-uniform} left). Because of the
higher concentration of light, the details of the triangulation are
more concentrated in the middle in the Gaussian case (middle) than in
the uniform case (right).

\begin{figure}
    \centering


    \includegraphics[width=.32\linewidth]{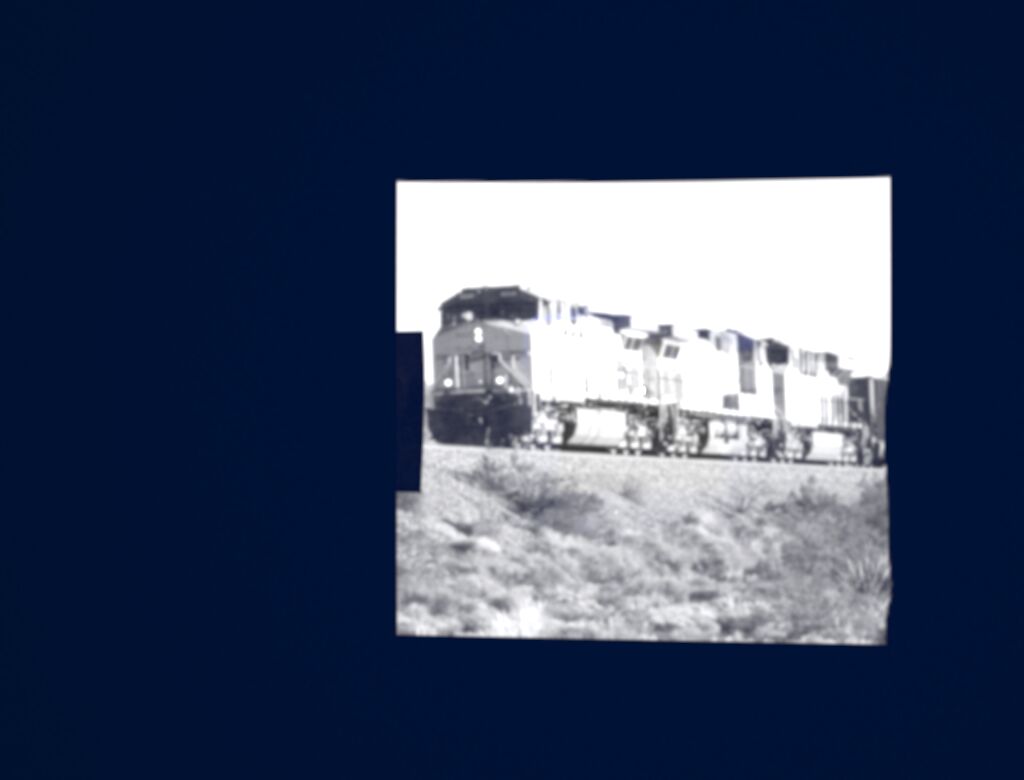}
    \includegraphics[width=.32\linewidth]{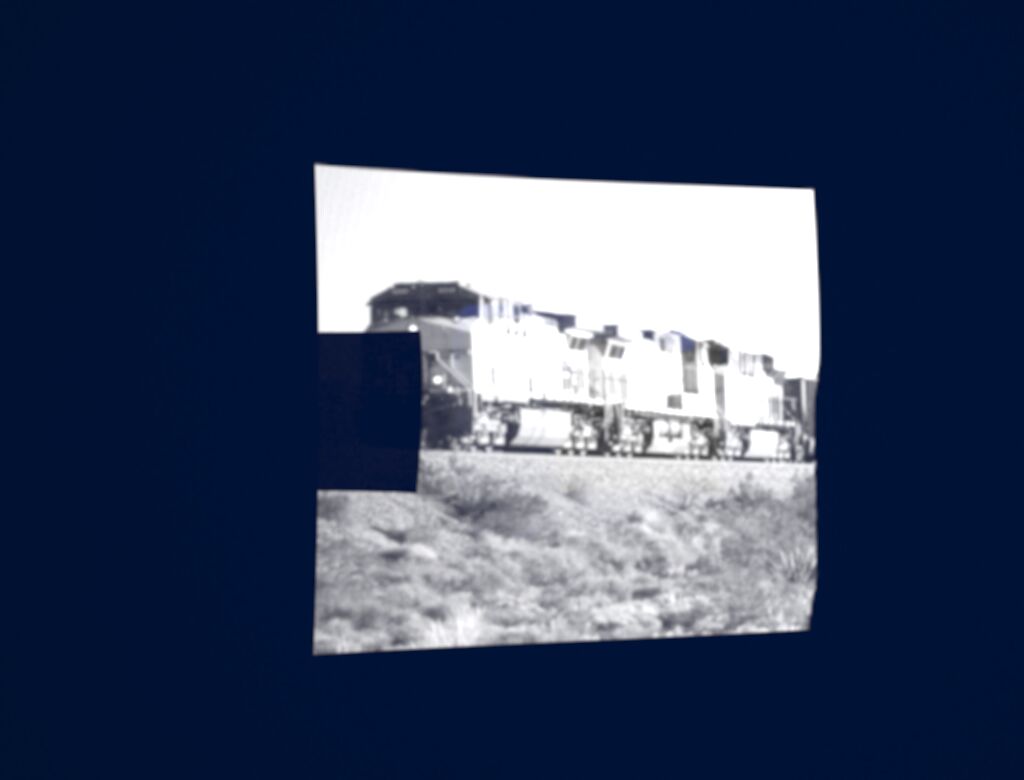}
    \includegraphics[width=.32\linewidth]{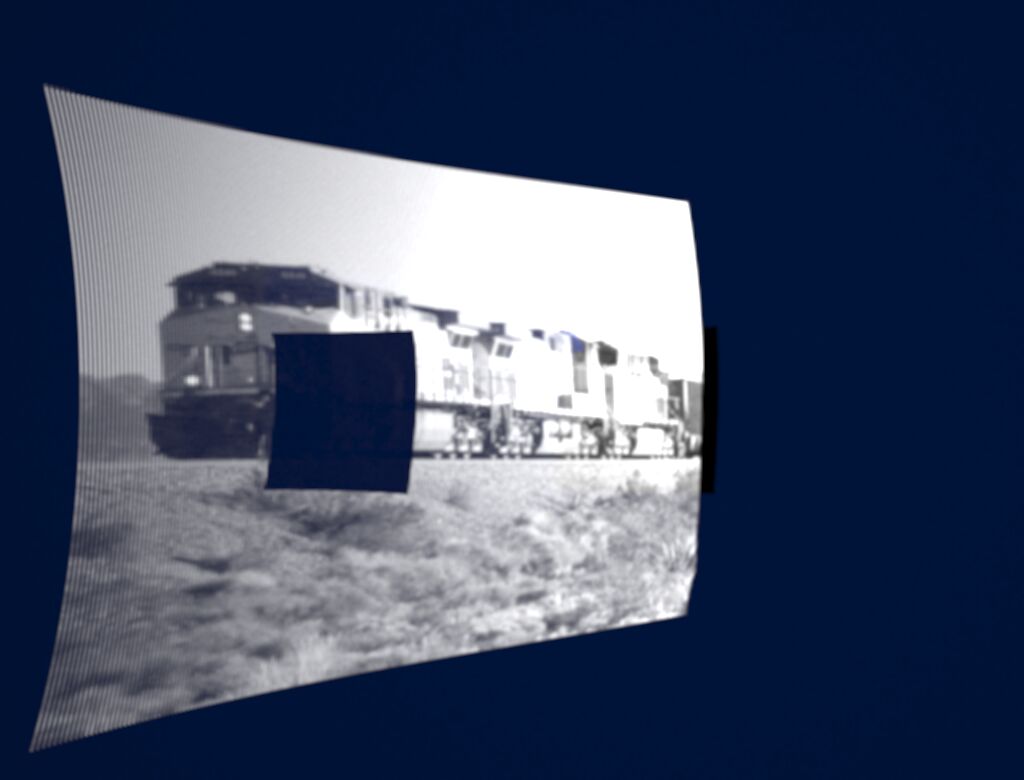}

    \caption{\jocelyn{\textbf{Stability under rotation of the lens.} LuxRender renderings in the \cslens setting for the
            \textsc{Train} target while rotating the lens with respect to the direction of the collimated light source ($0^{\circ}$ / $5^{\circ}$ / $15^{\circ}$).}
    }
    \label{fig:cs-lens-cameraman-angles}
\end{figure}

\paragraph{Convex / concave optical components.}
As shown in Section \ref{sec:LEC}, for each problem, one can choose
between two different parameterizations. For instance, for the \cslens
problem, one can build a lens which is either concave or convex, see
Figure~\ref{fig:res-cs_lens-concave-convex} for an illustration of
these differences.
\jocelyn{Note that in the \pslens setting (which corresponds to the last row of
Figure~\ref{fig:diagrams-problems} and Figure~\ref{fig:res-ps_lens}), the light
source is not supported on the full hemisphere $ \Sph^2_+ $ but instead on a
smaller part of it. Indeed, choosing a smaller support for $ \mu $ enforces that
$ \Ref_T $ is a graph above the plane instead of the hemisphere and thus avoids
potential inter-refractions. \jocelynbis{Furthermore, since we 
    parametrize the lens as a union of ellipsoids, it is neither convex nor
    concave.}} As for all figures, we have performed no post-processing on $\Ref_T$ in order to
emphasize the benefit of designing convex or concave optical components
(convexity is a form of regularity).
\jocelyn{One also observes that when the lens is rotated with respect to the light source
(Figure~\ref{fig:cs-lens-cameraman-angles} and first row of
Figure~\ref{fig:milling}), or when the target screen is not at the right
distance (Figure~\ref{fig:photo-depths}), the image is deformed in a monotonic
and regular way. We believe this is due to the monotonicity properties of
optimal transport and to the \jocelynbis{convexity/concavity} properties of the optical components. 
}

\paragraph{Comparison with previous work.}

\begin{figure}
    \centering
    \includegraphics[width=.8\linewidth]{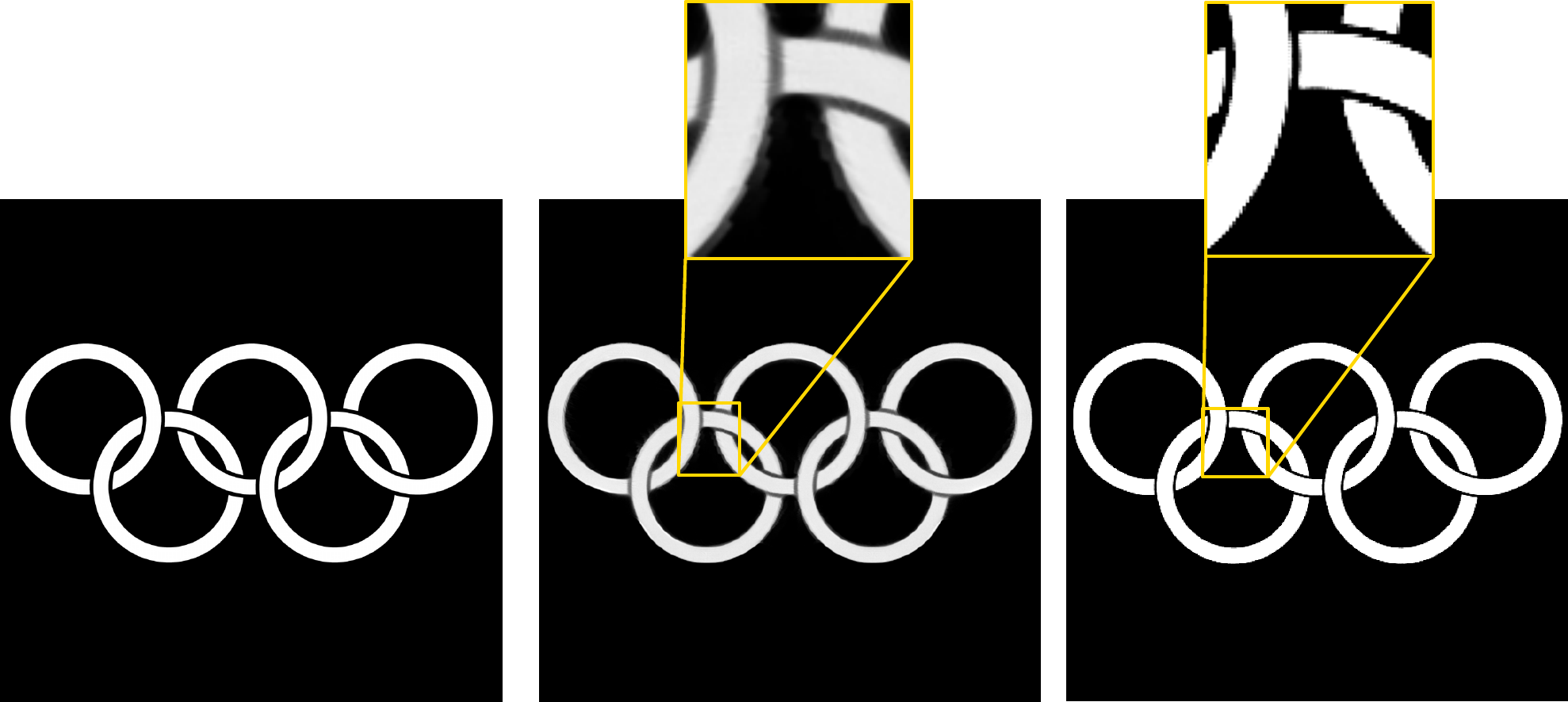}
    \includegraphics[width=.8\linewidth]{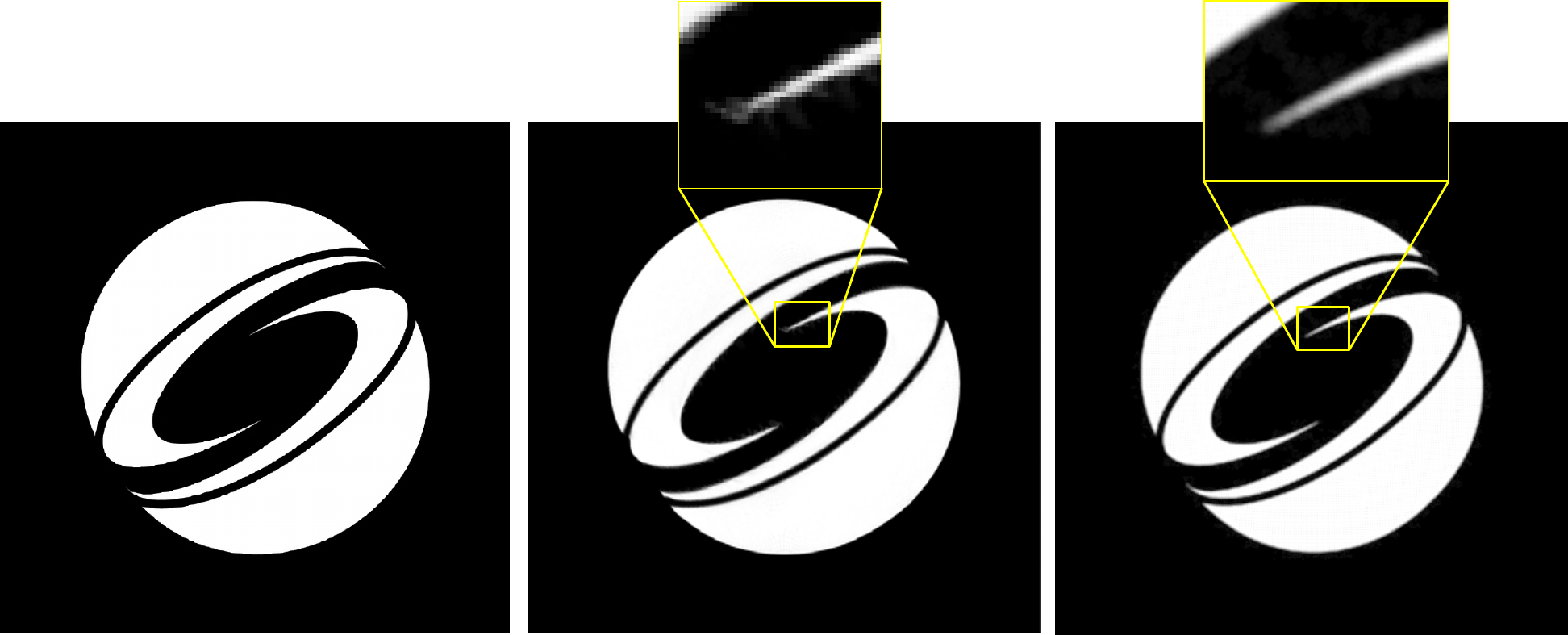}

    \includegraphics[width=.4\linewidth]{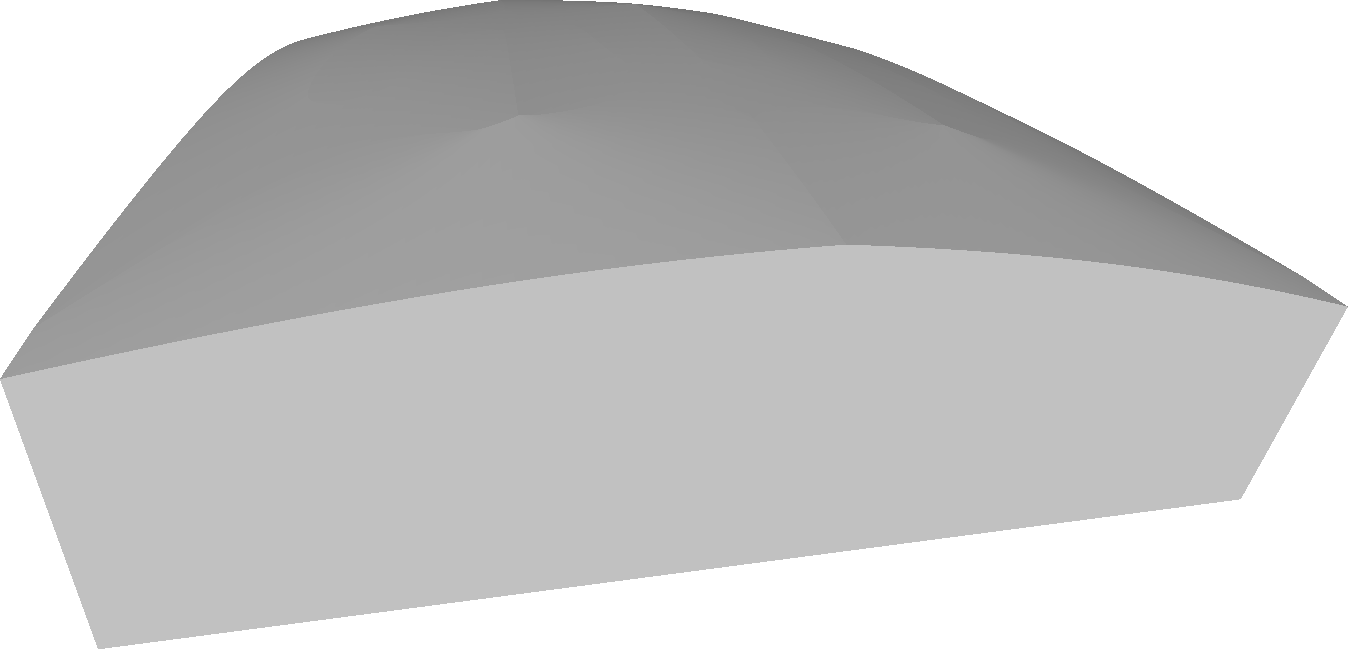}
    \includegraphics[width=.4\linewidth]{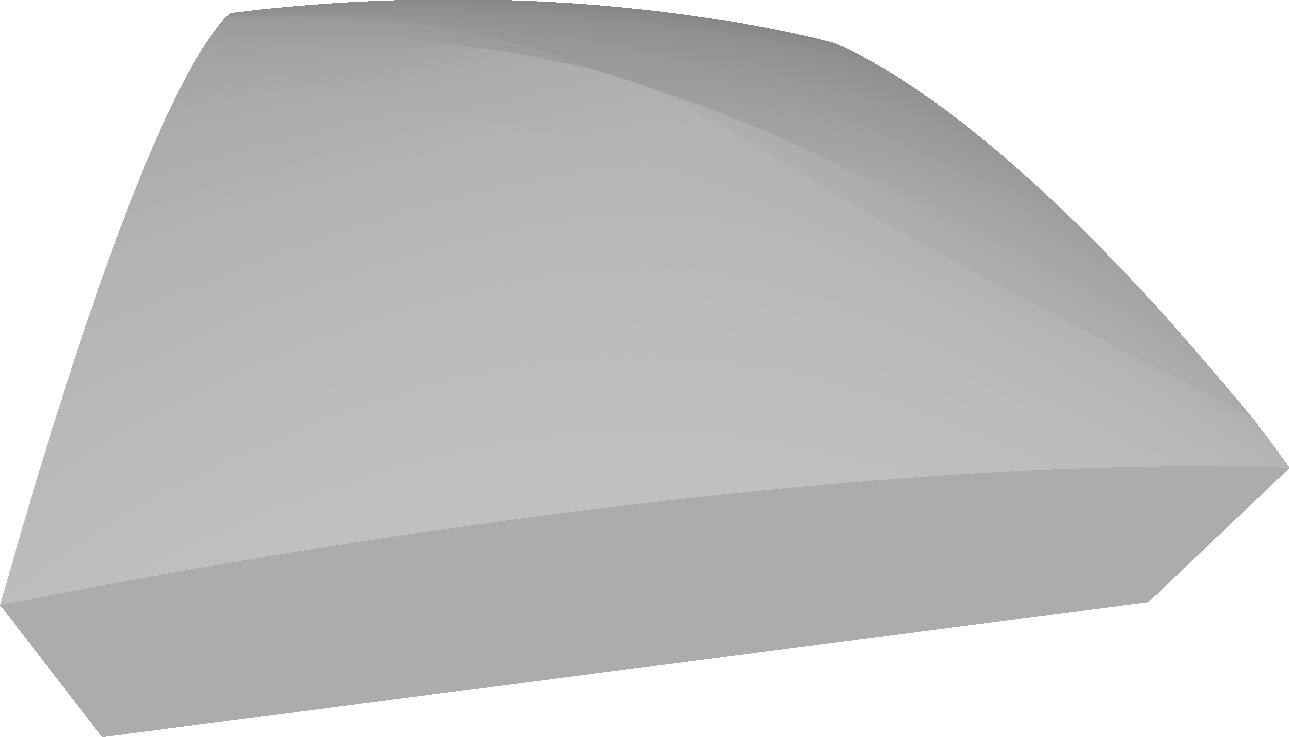}




    \caption{\textbf{Comparison with \protect\cite{schwartzburg2014high}} \textit{From left to right:} target distribution; images obtained
        by \protect\cite{schwartzburg2014high} and taken from their article; our forward
        simulation using \jocelyn{LuxRender}.
        \jocelyn{Last row: meshes of the two corresponding convex lenses:
            \textsc{Rings} (left) and \textsc{Siggraph} (right).}
    }
    \label{fig:comp-pauly}
\end{figure}

Figure \ref{fig:comp-pauly} compares
the state of the art results obtained by \citeauthor{schwartzburg2014high}
\shortcite{schwartzburg2014high} (second
column) and the LuxRender renderings obtained by our method (third column)
on two target distributions for the  \cslens case with a
collimated uniform light source \jocelyn{in the near-field setting}. \jocelyn{Although the results are comparable,} one can notice, in the second column, the presence of small artifacts between the black and white regions, for instance around the rings (notably in the center). The contrast is more accurate with our convex lenses.


\paragraph{Application to pillows}
This problem consists in decomposing the optical component (mirror or
lens) into several smaller optical components that are called
\textit{pillows}, as illustrated in Figure~\ref{fig:teaser}, and are
widely used in car headlight design. Each pillow independently
satisfies a non-imaging problem with the same target light, but
with a different source (since it receives only a portion of the
light). Hence, the optical component made with all the pillows
glued together is more reliable and allows for example to reduce the
artifacts due to small occluders. Indeed if one object is in front of
one or more lenses, the quality of the refracted image decreases but
the image can still be recognized. Using pillows also gives some
flexibility to the designer to improve the appearance and the volume
occupied by the component. An example with 9 pillows can be found in
Figure~\ref{fig:pillows}, and the effect of a small occluder.
In practice, in order to avoid a shift between the nine simulated
images, we solve the near-field problem
(see \jocelyn{Section~\ref{sec:near-field})}.

\jocelyn{
\begin{figure}
    \centering
    \includegraphics[height=3cm]{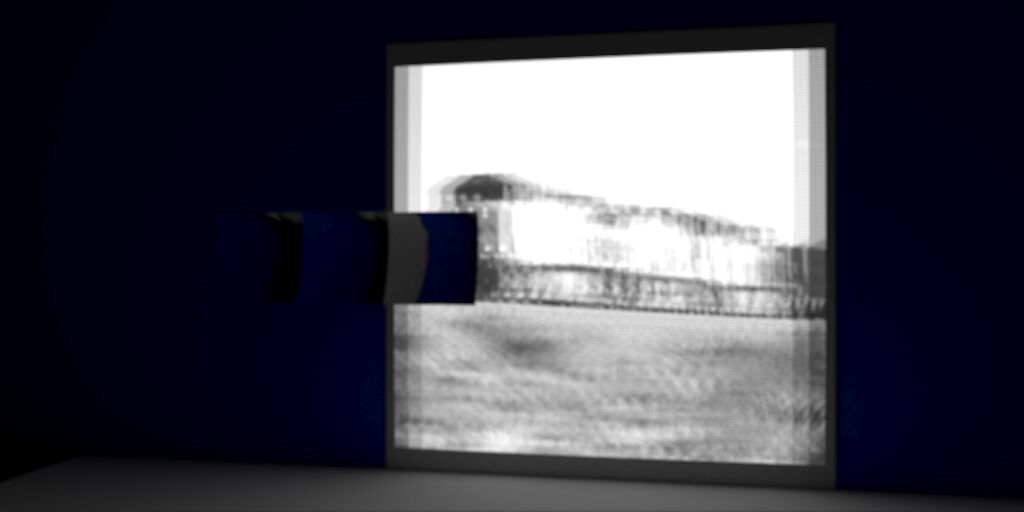}
    \includegraphics[height=3cm]{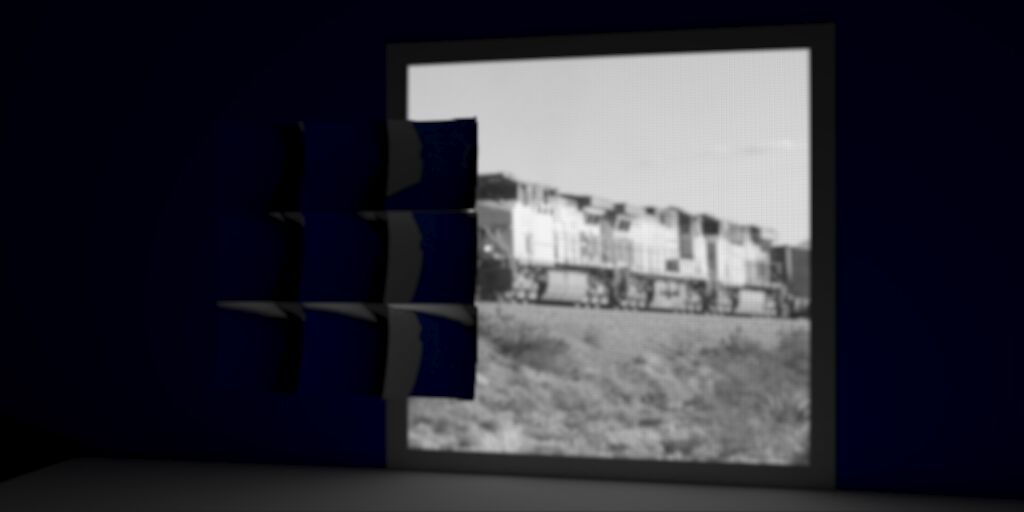}
    \includegraphics[height=3cm]{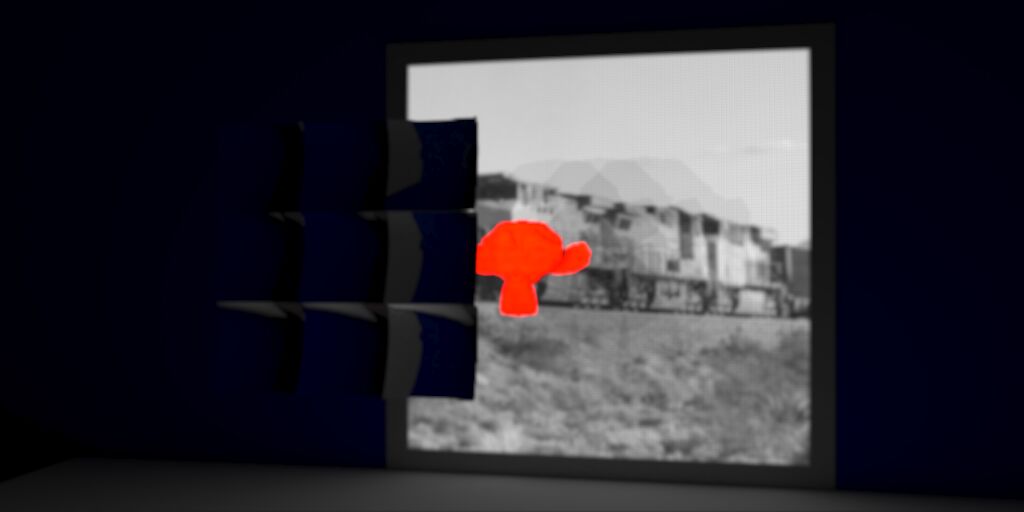}
    \caption{\textbf{Pillows and differences between FF and NF.}
        \textit{Top}: The lens is composed of three pillows that
        solve the FF problem. \jocelynbis{Note} in the last image the shift between the three projected images.
        \textit{Middle}: A lens composed of nine pillows (each of them solving the NF problem) that refracts a uniform
        collimated light source;
        \textit{Bottom}: The same lens with an obstacle in red. }
    \label{fig:pillows}
\end{figure}
}

%

\jocelyn{\paragraph{Application to color images}

Using pillows, we can also target color images. Indeed, we can build one component for each of the Red, Green and
Blue channels of an image. If we then place three lights (red, green and blue) in front
of each component, using Algorithm~\ref{algo:near-field} the 3 images will be perfectly align
and thus produce the original color image, see the first image of Figure~\ref{fig:teaser}.
}



\section{Physical prototypes}

\begin{figure*}
  \begin{minipage}{.49\linewidth}
\begin{center}
    \includegraphics[width=.99\linewidth]{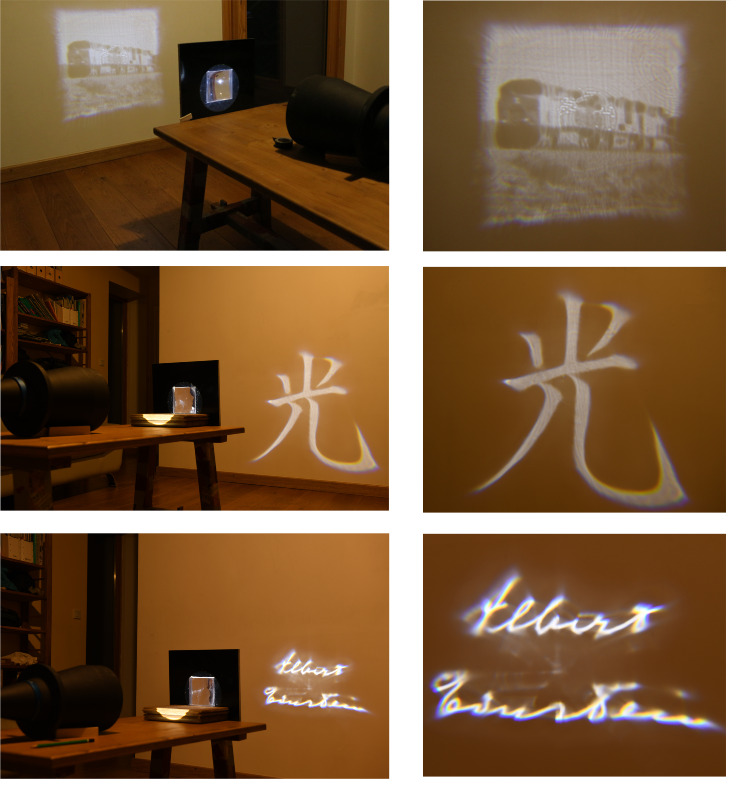}




    \caption{\textbf{Fabricated lenses for a collimated light source.} From left
        to right: experimental setup, zoom on the target screen. From top to
        bottom: \textsc{Train}, \textsc{Hikari}, \textsc{Einstein's signature}
        targets. Images are focused on a screen at 2
        meters for the first two rows and 1 meter for the last one.}
    \label{fig:photo-scenes}
    \end{center}
  \end{minipage} \hfill
  \begin{minipage}{.49\linewidth}
    \begin{center}
    

        \includegraphics[width=.67\linewidth]{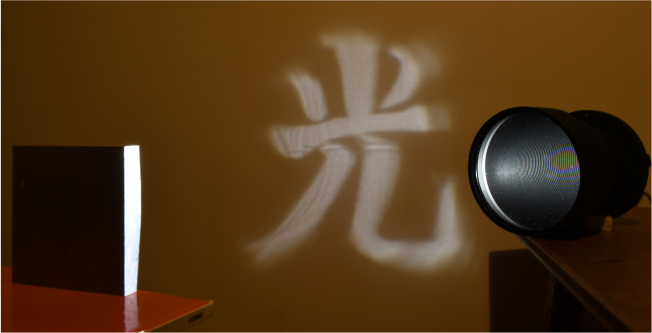}

    \caption{\textbf{Fabricated mirror for a collimated light source.}
      \textsc{Hikari} target. 
      \label{fig:photo-scenes-mirrors} \vspace{.3cm}}

    \includegraphics[width=.85\linewidth]{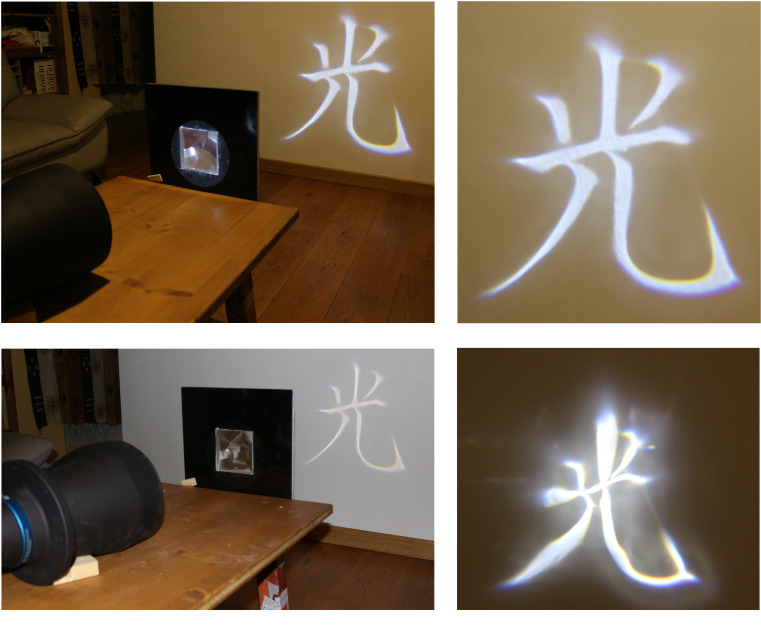}




    \caption{\textbf{Stability with respect to the depth of the focus plane.}
        The lens of \textsc{Hikari} is designed to focus at a distance of 2
        meters. The target screen is at different depths, top: 1.5 meters;
        bottom: 1m (left), 50cm (right).}
    \label{fig:photo-depths}

    \end{center}
    \end{minipage}
    \end{figure*}




\begin{figure}
    \centering

    \includegraphics[width=.9\linewidth]{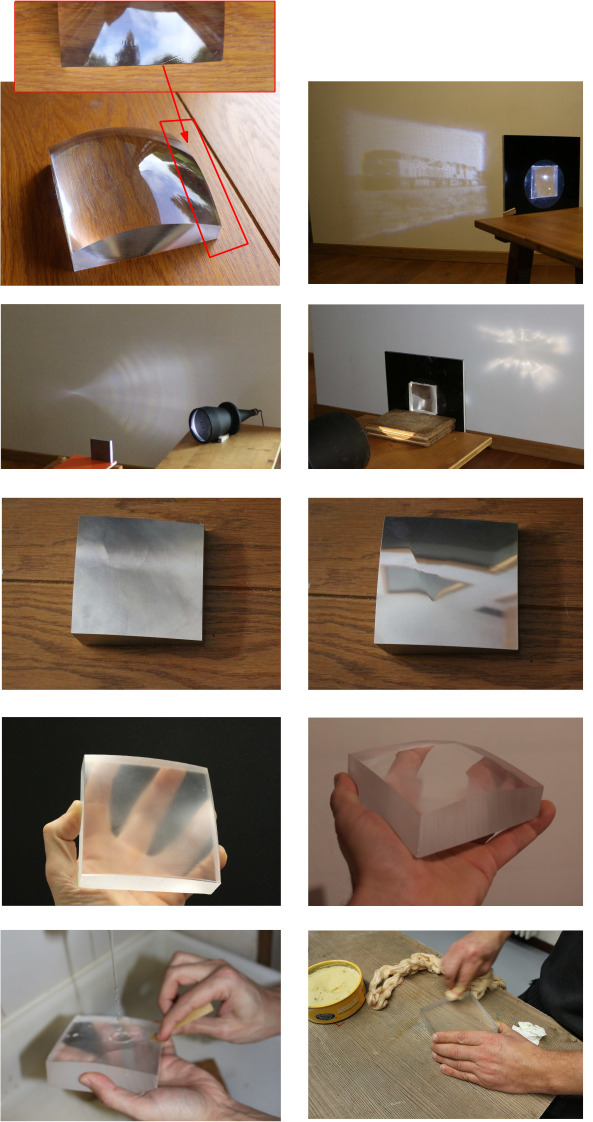}




    \caption{\textbf{Fabrication process.} \textit{First row:} lens for the
        \textsc{Train} target with a zoom on milling errors (left). The lens is rotated (20 degrees) around
        the axis of the light source (right).
        \textit{Second row:} Collimated light projected after reflection or
        refraction on the screen (rough mirror and rough \textsc{Einstein's signature})
        \textit{Third row:} Rough mirror / sandpapered and polished mirror.     
        \textit{Fourth row:} Rough lens / sandpapered and  polished lens. 
        \textit{Fifth row:} Sandpapering with water / Polishing by hand.
        }
    \label{fig:milling}
\end{figure}

We built three lenses (see Figure~\ref{fig:photo-scenes}) and two mirrors (see figures~\ref{fig:photo-scenes-mirrors} and \ref{fig:teaser}) corresponding to collimated light sources. The lenses are fabricated in PMMA (whose index of refraction is $1.49 $) and the mirrors in aluminum. 
All the lenses and mirrors have size $100mm\times100mm$ and were milled in one
pass on 3-axis CNC machines after milling the blank. For the Hikari lens, we choose to focalize the target image on a wall at $2$ meters and the target is a square of size $600mm\times600mm$. For the Train and Einstein signature lenses and the two mirrors, we choose to focalize the target image on a wall at $1$ meter and the target is a square of size $300mm\times300mm$. The five components were milled in one pass on the \emph{DMG-DC100V} machine with a $10mm$ \jocelyn{radius end-mill}. 

The milling process is very sensitive. For the lenses, the end-mill is a super finishing ball mill D10, 3 teeth and is following a \jocelyn{concentric spiral} trajectory. For the mirrors the end-mill is a PCD ball mill hooped D10, 2 teeth and is following a parallel scanning trajectory.
We observe that the precision of the milling is not accurate enough: when the collimated light source is traversing a lens or reflected by a mirror with no sandpapering and polishing, the light is dispersed and we do not recognize the target (see Figure~\ref{fig:milling}, second row). We had to sandpaper them by hand before polishing them with a  polishing paste. This clearly damages the
lens surface: there is a tradeoff between removing the artifacts due to the
milling and smoothing too much the surface (see Figure~\ref{fig:milling} rows
3-5), thus damaging its refractive properties. \jocelynbis{Note} that thanks to the convexity property (see Figure~\ref{fig:milling} row 4), the lens surface is quite regular and is more robust to sandpapering.

We can also observe some artifacts in the milling process. For instance, some
corrugations are present in the lens (Figure~\ref{fig:milling} first row) and induce some artifacts in the projected image (Figure~\ref{fig:photo-scenes}, first row). We observe that although the image are very contrasted, the projected image are very accurate. The boundary of the target is often slightly blurred and this is due to the boundary of the lens or mirror where the milling was less good. Our model do not take into account the different wavelengths of the white color and we observe on the boundary of the projected images a small chromatic aberration (the boundary is slightly blue).


\section{Conclusion and perspectives}

\jocelyn{
We presented a general framework for eight different optical component
design problems satisfying light energy constraints. We proposed an
efficient algorithm able to solve them \jocelyn{whether the target is at
    infinity (far-field) or at a finite distance (near-field).} 
The main limitation of the approach is the fact that we only deal with
ideal light sources (a light bulb is for instance neither collimated nor punctual).
Another limitation is that we do not account for self
shadowing and internal reflections (although, this is not a problem in
the situations we have encountered).
In the future, we also want to try fabricating physical
prototypes when the source is punctual. This problem becomes more difficult for the lenses since 
we also have to mill the two sides of the lens, and in particular the inner sphere. 
We also believe that the robustness and versatility of the
proposed approach can make it a useful component for the design of
heuristics able to deal with extended light sources and in computer
graphics for caustic design.
}

\paragraph{Acknowledgements}
We would like to thank Alain Di Donato from the \textit{GINOVA} platform, \textit{S.MART DS} department for
building the physical prototypes, as well as the GMP department of the Institute
of Technology of Aix-en-Provence in France, and Andr\'e Lieutier and Jimmy
Cresson from Dassault-Syst\`emes for helping with the fabrication process, and Philippe Halot for
taking the pictures. This
work has been partially supported by the LabEx PERSYVAL-Lab
(ANR-11-LABX-0025-01) and by ANR project
MAGA (ANR-16-CE40-0014 MAGA).

\bibliographystyle{apalike}
\bibliography{reflector}

\end{document}